\documentclass[twocolumn,showpacs,amsmath,amssymb]{revtex4-1}
\usepackage{bm}
\usepackage[T1]{fontenc}
\usepackage{hyperref}
\usepackage{tabularx}
\input{epsf}

\usepackage{graphicx}
\usepackage{epstopdf}
\usepackage{array}
\usepackage{longtable}
\usepackage{rotating,booktabs}
\usepackage{booktabs,threeparttable}
\usepackage{bm}
\usepackage{float}
\usepackage{amsmath}
\usepackage{gensymb}
\usepackage{multirow}
\usepackage{epsfig}
\usepackage{color}

\begin{document}
 \title{Determination of nuclear quadrupole moments for $^{25}$Mg, $^{87}$Sr, and $^{135,137}$Ba via configuration-interaction combined with a coupled-cluster approach}
\author{Yong-Bo Tang}
\affiliation {College of Engineering Physics, Shenzhen Technology University, Shenzhen, 518118, China}
\email{tangyongbo@sztu.edu.cn}
\date{\today}

\begin{abstract}
Using the configuration-interaction plus coupled-cluster approach, we calculate the electric-field gradients $q$ for the low-lying states of alkaline-earth atoms, including magnesium (Mg), strontium (Sr), and barium (Ba). These low-lying states specifically include the $3s3p~^3\!P_{1,2}$ states of Mg; the $5s4d~^1\!D_{2}$ and $5s5p~^3\!P_{1,2}$ states of Sr; as well as the $6s5d~^3\!D_{1,2,3}$, $6s5d~^1\!D_{2}$, and $6s6p~^1\!P_{1}$ states of Ba. By combining the measured electric quadrupole hyperfine-structure constants of these states, we accurately determine the nuclear quadrupole moments of $^{25}$Mg, $^{87}$Sr, and $^{135,137}$Ba. These results are compared with the available data. The comparison shows that our nuclear quadrupole moment of $^{25}$Mg is in perfect agreement with the result from the mesonic X-ray experiment. However, there are approximately 10\% and 4\% differences between our results and the currently adopted values [Pyykk$\rm \ddot{o}$, Mol. Phys. 116, 1328(2018)] for the nuclear quadrupole moments of $^{87}$Sr and $^{135,137}$Ba respectively. Moreover, we also calculate the magnetic dipole hyperfine-structure constants of these states, and the calculated results exhibit good agreement with the measured data.
 \end{abstract}

\maketitle
\section{Introduction}\label{sec1}

The nuclear electric quadrupole moment $Q$ is an important parameter offering fundamental information on nuclear structure symmetry. Accurate knowledge of this parameter is important in multiple scientific fields like atomic, nuclear, and condensed matter physics, as well as chemistry and biology~\cite{QiPRA2023,QiPRR2025,GuanPRAL2025,Heyde2011RMP,Campbell2016PRNP,Sinitsyn2012PRL,Pekka2018MP,WongCPL2006}. An effective way to determine nuclear quadrupole moments is to integrate the experimentally-measured electric quadrupole hyperfine-structure constants (\(B=234.9648867qQ\), in MHz), with the theoretically-calculated electric-field gradients $q$ (in $a.u.$). This method is independent of nuclear models and is one of the most accurate methods for extracting the nuclear quadrupole moments of unstable nuclei and heavy nuclei~\cite{Sahoo2013pra,Bi2018pra,Li2021pra,Porsev2021prl,Papoulia2021pra,Skripnikov2021prc,Skripnikov2024prc,Lu2024pra,Zhang2025pra,Tang2025pra}. Its accuracy in extracting the nuclear quadrupole moment depends on calculated electric-field gradient as hyperfine spectra measured by spectral technology are high nowadays. Precise electric-field gradient calculation requires considering relativistic and electron correlation effects. Since the relativistic effect can be easily dealt with by solving the Dirac-Fock equation, electron correlation is the decisive factor for accurate electric-field gradients.

Recently, we developed a relativistic hybrid-method-based code for accurately calculating divalent atomic system properties. The hybrid method integrates the advantages of the configuration interaction method and the coupled-cluster method, and can simultaneously account for core-core, core-valence, and valence-valence correlations. We used it to calculate electric-field gradients of low-lying states in \(^{43}\mathrm{Ca}\) and got an accurate nuclear quadrupole moment of \(^{43}\mathrm{Ca}\)~\cite{Tang2025pra}. In the present work, we apply the same method to obtain nuclear quadrupole moments of \(^{25}\mathrm{Mg}\), \(^{87}\mathrm{Sr}\), and \(^{135,137}\mathrm{Ba}\).

In the latest nuclear quadrupole moment literature~\cite{Pekka2018MP}, the recommended nuclear quadrupole moment of \(^{25}\text{Mg}\) is 0.1994(20)~b, as derived from the hyperfine spectrum of the \(3s3p\ ^{3}P_{2}\) state~\cite{Sundholm1991npa}. This value is in good agreement with the 0.201(3)~b obtained from mesonic X-ray experiments~\cite{Weber1982npa}. Regarding \(^{87}\text{Sr}\), the recommended value in Ref.~\cite{Pekka2018MP} is 0.305(2)~b, which is determined from the hyperfine spectrum of the \(\text{Sr}^+\) \(4d_{5/2}\) state~\cite{Barwood2003pra,Sahoo2006pra}. In contrast, Lu \textit{et al.} reported a value of 0.328(4)~b from the \(5s5p\ ^{3}P_{1,2}\) states of the neutral Sr atom \cite{Lu2019pra}, approximately 8\% larger than the recommended value~\cite{Pekka2018MP}. For \(^{137}\text{Ba}\), the recommended value in Ref.~\cite{Pekka2018MP} is 0.236(3)~b, extracted from the \(\text{Ba}^+\) \(5d_{3/2}\) and \(5d_{5/2}\) states~\cite{Sahoo2013pra}. This value is about 4\% smaller than those obtained from the \(6p_{3/2}\) state~\cite{Silverans1986pra,Wendt1988ZPA}. Reported \(^{25}\text{Mg}\) nuclear quadrupole moment values are relatively consistent, while there are notable discrepancies in \(^{87}\text{Sr}\) and \(^{137}\text{Ba}\) values across studies. Thus, a comprehensive and in-depth reinvestigation of \(^{87}\text{Sr}\) and \(^{137}\text{Ba}\) nuclear quadrupole moments is necessary.

The nuclear quadrupole moment can be obtained from hyperfine spectra computations and measurements of neutral atoms or ions. Actually, the hyperfine spectra of \(3s3p~^3\!P_{1,2}\) in \(^{25}\mathrm{Mg}\)~\cite{LurioPR1961}, \(5s5p~^3\!P_{1,2}\) and \(5s4d~^1\!D_{2}\) in \(^{87}\mathrm{Sr}\)~\cite{GrundevikZPA1983,PutlitzZP1963,HeiderPRA1977}, and \(6s6p~^1\!P_{1}\), \(6s5d~^3\!D_{1,2,3}\), \(6s5d~^1\!D_{2}\) in \(^{135,137}\mathrm{Ba}\)~\cite{GustavssonZPA1979,SchmellingPRA1974,vanCJP1995} have been precisely measured, and their electric quadrupole hyperfine-structure constants have been determined. These accurately-determined electric quadrupole hyperfine-structure constants are very suitable for extracting the nuclear quadrupole moments of \(^{25}\mathrm{Mg}\), \(^{87}\mathrm{Sr}\), and \(^{135,137}\mathrm{Ba}\) if the electric-field gradients of these states can be calculated with high precision.
The present work aims to systematically calculate electric-field gradients of these low-lying states of neutral Mg, Sr, and Ba atoms. Energies and magnetic-dipole hyperfine-structure constants of these low-lying states are also calculated and compared with available experimental and theoretical data to assess the precision of our results.

The structure of this paper is as follows. In Sec.~\ref{theory}, we present the formulations of the hyperfine interaction and the theoretical method. Sec.~\ref{results} is dedicated to presenting the numerical results, along with in-depth discussions. Additionally, in this section, we also compare our results with the available experimental and theoretical data. Finally, Sec.~\ref{conclusions} provides a summary. Unless explicitly specified otherwise, atomic units are employed throughout the paper.

\section{Theoretical equations}\label{theory}

\subsection{Hyperfine interaction}
The hyperfine interaction is characterized by the interaction between electrons and the electromagnetic multipole moments of the nucleus. The Hamiltonian can be formulated as follows:
\begin{eqnarray}\label{eq1}
	H_{\rm {HFI}}=\sum _{k}{\textbf{\rm{T}}^{(k)}}\cdot \textbf{\rm{N}}^{(k)},
\end{eqnarray}
where $\textbf{\rm{T}}$ and $\textbf{\rm{N}}$ respectively denote the spherical tensor operators of rank $k$ $(k > 0)$ in the electronic and nuclear coordinate spaces. According to the parity and angular selection rules, $k$ must be an even number for electric moments and an odd number for magnetic moments. Specifically, $k = 1$, $k = 2$, and $k = 3$ correspond to the magnetic dipole, electric quadrupole, and magnetic octupole hyperfine interactions, respectively. In this work, we only take into account the magnetic dipole and electric quadrupole hyperfine interactions.
\begin{widetext}
The matrix element between two hyperfine states can be computed as:
\begin{small}
  \begin{equation}
  \begin{aligned}
\left\langle \gamma^{\prime} I J^{\prime} F^{\prime} M_{F\prime}^{\prime}\left|H_{\mathrm{HFI}}\right| \gamma I J F M_{F}\right\rangle
=\delta_{F^{\prime} F} \delta_{M_{F\prime}^{\prime} M_{F}}\times (-1)^{I+J'+F} \sum_{k}\left\{\begin{array}{lll}
			  F & J & I \\
			   k & I & J^{\prime}
			    \end{array}\right\}\left\langle\gamma^{\prime} J^{\prime}\left\|T^{(k)}\right\| \gamma J\right\rangle\left\langle I \left\|N^{(k)}\right\| I\right\rangle,
		  \end{aligned}
\end{equation}
\end{small}
where $I$, $J$, and $F$ represent the nuclear spin, total angular momentum of the electrons, and the total angular momentum, respectively. The state $|\gamma IJ F M_{F}\rangle$ is the hyperfine state constructed by coupling a nuclear eigenstate $|IM_{I}\rangle$ with an atomic eigenstate $|\gamma JM_{J}\rangle$, where $\gamma$ represents the remaining electronic quantum numbers.
The first-order correction $\Delta{E_{F}^{(1)}}$ of the hyperfine interaction to the energy can be defined as:

  \begin{equation}
    \label{e1}
    \begin{aligned}
    \Delta{E_{F}^{(1)}}=(-1)^{I + J+ F}\sum_{k}\left\{\begin{array}{lll}
    F & J & I \\
    k & I & J
    \end{array}\right\}\left\langle\gamma J\left\|T^{(k)}\right\| \gamma J\right\rangle\left\langle I\left\|N^{(k)}\right\| I\right\rangle.
    \end{aligned}
    \end{equation}

The nuclear matrix elements are expressed in terms of conventional nuclear moments as:

\begin{equation}
      \left\langle I\left\|N^{(1)}\right\| I\right\rangle=\mu\sqrt{\frac{(I + 1)(2I + 1)}{I}}, \\
\end{equation}
\begin{equation}
      \left\langle I\left\|N^{(2)}\right\| I\right\rangle=\frac{Q}{2}\sqrt{\frac{(I + 1)(2I + 1)(2I + 3)}{I(2I - 1)}}, \\
\end{equation}
where $\mu$ is the nuclear magnetic dipole moment, and $Q$ is the nuclear electric quadrupole moment. When restricted to $k\leq 2$, $\Delta{E_{F}^{(1)}}$ can be parameterized as:
\begin{align}
\Delta{E_F^{(1)}}=&\frac{A}{2}K+\frac{B}{2}\frac{3K(K + 1)-4I(I + 1)J(J + 1)}{2I(2I - 1)2J(2J - 1)},
\end{align}
where $K = F(F + 1)-I(I + 1)-J(J + 1)$, and $A$ and $B$ are the magnetic dipole and electric quadrupole hyperfine-structure (hfs) constants, which are defined as:

\begin{align}
A=\frac{\mu}{I}\frac{\langle\gamma J\|T^{(1)}\|\gamma J\rangle}{\sqrt{J(J + 1)(2J + 1)}},
\end{align}

and
\begin{align}
B=2Q\bigg[\frac{2J(2J - 1)}{(2J + 1)(2J + 2)(2J + 3)}\bigg]^{1/2}\langle\gamma J\|T^{(2)}\|\gamma J\rangle,
\end{align}
respectively. Here, $T^{(k)}=\sum_{i}{t^{(k)}(\textbf{r}_{i})}$. The single-particle reduced matrix elements of the operators $t^{(1)}$ and $t^{(2)}$ are given by:

\begin{align}
\begin{cases}
\langle{\kappa_{a}}\|t^{(1)}\|\kappa_{b}\rangle=-(\kappa_{a}+\kappa_{b})\langle-\kappa_{a}\|C^{(1)}\|\kappa_{b}\rangle
\int_{0}^{\infty}{\frac{f_{a}(r)g_{b}(r)+f_{b}(r)g_{a}(r)}{r^2}\times{F^{(1)}(r)}dr}\\
\langle{\kappa_{a}}\|t^{(2)}\|\kappa_{b}\rangle=-\langle\kappa_{a}\|C^{(2)}\|\kappa_{b}\rangle\int_{0}^{\infty}{\frac{f_{a}(r)f_{b}(r)+g_{a}(r)g_{b}(r)}{r^3}\times{F^{(2)}(r)}dr}
\end{cases},
\end{align}
and
 \begin{align}\label{18}
    \langle\kappa_a\parallel C^{(k)}\parallel\kappa_b\rangle&=(-1)^{j_a+\frac{1}{2}}\sqrt{(2j_a + 1)(2j_b + 1)}
    \times
    \left(\begin{array}{ccc}
        j_a&j_b&k\\
        -\frac{1}{2}&\frac{1}{2}&0\\
    \end{array}\right)\Pi(\ell_a,L,\ell_b),
\end{align}
\end{widetext}
where \(\Pi(\ell_a,k,\ell_b)=1\) when \(\ell_a+k+\ell_b\) is even; otherwise, \(\Pi(\ell_a,k,\ell_b)=0\). The relativistic angular-momentum quantum number \(\kappa=\ell(\ell + 1)-j(j + 1)-\frac{1}{4}\). $f$ and $g$ are the large and small radial components of the Dirac wave function, respectively. The nuclear distribution function $F^{(k)}(r)$ is defined as:
\begin{align}
	F^{(k)}(r) =\begin{cases}
		(\frac{r}{R_{N}})^{2k + 1},&r\leq{R_{N}}\\
		1,&r > R_{N}
	\end{cases},
\end{align}
where $R_{N}=\sqrt{5/3}\langle{r^{2}}\rangle^{1/2}$ is the radius of the sphere, and $\langle{r^{2}}\rangle^{1/2}$ is the charge root-mean-square radius of the nucleus. In the present work, the nucleus is modeled as a uniformly magnetized and charged sphere.

Based on the above definitions of the hfs constants, the nuclear electric quadrupole moment $Q$ (in b) can be extracted from the experimental values of the hfs constant $B$ using the formula:
\begin{align}
Q=\frac{B}{234.9648867q},
\end{align}
where the hfs constant $B$ is in the unit of MHz, and the electric-field gradient $q$ is defined as
\begin{small}
\begin{align}\label{eq9}
{q}=2\bigg[\frac{J(2J - 1)}{(2J + 1)(J + 1)(2J + 3)}\bigg]^{1/2}\langle\gamma J\|T^{(2)}\|\gamma J\rangle.
\end{align}
\end{small}
and is expressed in atomic units.

\subsection{Configuration interaction plus coupled-cluster method}
In the many-electron atomic system, electron-electron correlation encompasses core-core, core-valence, and valence-valence correlations. All three types of correlation effects are of paramount importance for the precise evaluation of hyperfine interaction properties.
In the present work, we adopt a hybrid approach that integrates the relativistic configuration interaction with coupled-cluster methods to comprehensively account for these three types of electron-electron correlations. Within the framework of this hybrid methodology, coupled-cluster calculation is employed to depict the core-core and core-valence correlations, while the valence-valence correlation is taken into consideration through a configuration interaction calculation. The relativistic configuration interaction plus a linearized version of coupled-cluster theory, called the RCI+all-order method, was first developed by Safronova \textit{et al.}~\cite{Safronova2009PRA}; subsequently, Dzuba independently devised a similar approach~\cite{Dzuba2014PRA}.
The concept of this hybrid approach can be traced back to the RCI+MBPT combination introduced by Dzuba and co-workers~\cite{Dzuba1996}.
The method utilized in our work is conceptually similar to theirs. However, we have independently developed the corresponding software package for this hybrid method, therefore there remain a few differences in the handling details compared to their methods~\cite{Tang2025pra}.

For alkaline-earth atoms, the equation governing the effective interaction can be formulated as follows:
\begin{equation}\label{7}
    \left(\sum_{i = 1}^{2}(H_{\rm DF}(r_{i})+\Sigma_{1}(r_{i}))+\frac{1}{r_{12}}+\Sigma_{2}(r_{12})\right)\vert\gamma{JM}\rangle = E\vert\gamma{JM}\rangle,
\end{equation}
where \(H_{\rm DF}\) represents the Dirac-Fock Hamiltonian, and \(\Sigma_{1}\) and \(\Sigma_{2}\) correspond to the one-body and two-body correlation potentials, respectively. In this work, we derive the one-body and two-body correlation potentials via second-order many-body perturbation (MBPT(2)) calculations and coupled-cluster calculations.  Here we give only the effective equation for the valence electrons; the core electrons are incorporated into the one- and two-body correlation potentials, while the interaction equations for the many-electron system are detailed in Ref.~\cite{Dzuba1996}.

The wave function \(\vert\gamma{JM}\rangle\) of the system is represented as a linear combination of configuration wave functions that share the same parity, angular momentum \(J\), and magnetic quantum number \(M\). Specifically,
\begin{equation}\label{10}
    \vert\gamma{JM}\rangle=\sum_{v\leq w}C_{vw}\vert{vw}\rangle,
\end{equation}
where \(C_{vw}\) denote the expansion coefficients. The configuration wave function is constructed from single-particle orbitals as follows:
\begin{equation}\label{11}
    \vert{vw}\rangle=\eta_{vw}\sum_{m_{v},m_{w}}\langle j_{v}m_{v},j_{w}m_{w}\vert JM\rangle a_{v}^{\dagger}a_{w}^{\dagger}\vert0\rangle.
\end{equation}
The symmetry factor \(\eta_{vw}\) is defined as:
\begin{equation}\label{12}
    \eta_{vw}=\begin{cases}
        \frac{\sqrt{2}}{2},& v = w\\
        1,& v\neq w
    \end{cases}.
\end{equation}
By substituting Eq.~(\ref{10}) into Eq.~(\ref{7}) and applying the variational principle, we can obtain a general eigenvalue equation. Solving this eigenvalue equation enables us to determine the energies and wave functions of the system.

Afterwards, one can compute the transition matrix elements of various operators by utilizing the obtained wave functions. During the calculation of these transition matrix elements, we take into account the random phase approximation (RPA), core Brueckner effects, structural radiation, and normalization corrections of all orders~\cite{Tang2017PRA,Blundell1989PRA,Dzuba1998JETP}.
For the convenience of description, later we will combine the core Brueckner effects, structural radiation, and normalization corrections and refer to them as the high-order (HO) correction beyond RPA. Moreover, we include the two-particle (TP) interaction up to the second order~\cite{Dzuba1998JETP,Safronova1999JPB,Savukov2004PRA}. A detailed account of the method and the corresponding formulas has been provided in a recent work~\cite{Tang2025pra}.

\subsection{Computation details}

\begin{table}
\caption{The parameters of the Gauss basis set, along with the basis spaces employed in the CC and CI calculations. $N$ denotes the number of Gauss basis set. $N_{\rm core}$ represents the quantity of atomic core orbitals. $N_{\rm valence}$ and $N_{\rm virtual}$ signify the numbers of valence orbitals and virtual orbitals in the CC
calculation, respectively. $N_{\rm CI}$ stands for the number of orbitals utilized to construct the configurations.}
\label{tab1}
\begin{ruledtabular}
\begin{tabular}{ccccccccccccc}
 &$s$&$p$&$d$&$f$&$g$&$h$&$i$&$k$\\
\hline
\multicolumn{8}{c}{Mg}\\
$\alpha(\times{10^{2}})$&0.085&0.085&0.085&0.26&8.6&8.6&8.6&\\
$\beta$&1.88&1.87&1.89&1.91&2.0&2.0&2.0&\\
N    &35&30&30&25&15&15&15&\\
N$_{\rm core}$    &1-2&1&0&0&0&0&0&\\
N$_{\rm valence}$    &3-6&2-5&1-2&0&0&0&0&\\
N$_{\rm virtual}$  &3-28&2-27&1-25&1-22&1-15&1-15&1-15&\\
N$_{\rm CI}$   &3-22&2-21&1-21&1-19&1-15&0&0&\\

\multicolumn{8}{c}{Sr}\\
$\alpha(\times{10^{2}})$&0.055&0.055&0.055&0.15&8.6&8.6&8.6&\\
$\beta$&1.85&1.84&1.85&1.91&2.0&2.0&2.0&\\
N    &40&40&40&30&25&25&25&\\
N$_{\rm core}$    &1-4&1-3&1&0&0&0&0&\\
N$_{\rm valence}$    &5-8&4-7&2-5&&0&0&0&\\
N$_{\rm virtual}$  &5-30&4-29&2-27&1-24&1-20&1-20&1-20&\\
N$_{\rm CI}$   &5-23&4-22&2-21&1-20&1-15&0&0&\\

\multicolumn{9}{c}{Ba}\\
$\alpha(\times{10^{2}})$&0.055&0.055&0.055&0.15&8.6&9.6&9.6&9.6\\
$\beta$&1.87&1.85&1.86&1.91&2.0&2.2&2.2&2.2\\
N    &45&45&45&35&30&25&25&25\\
N$_{\rm core}$    &1-5&1-4&1-2&0&0&0&0&0\\
N$_{\rm valence}$ &6-9&5-8&3-6&0&0&0&0\\
N$_{\rm virtual}$  &6-30&5-29&3-27&1-24&1-21&1-20&1-20&1-20\\
N$_{\rm CI}$   &6-24&5-23&3-23&1-20&1-15&1-15&0&0\\
\end{tabular}
\end{ruledtabular}
\end{table}

In the present work, we use a finite basis set consisting of even-tempered Gaussian-type functions to expand the large and small components of the Dirac wave functions as in Refs.~\cite{Chaudhuri1999pra,Tang2025pra}. The Gaussian-type function has the form:
 \begin{align}
G_{i,\kappa}={\aleph}_{i}r^{n_{\kappa}}e^{-\alpha_{i}{r^2}},
\end{align}
where $\aleph_{i}$ is the normalization factor, $n_{\kappa}=\ell+1$, and $\alpha_{i}=\alpha\beta^{i-1}$.

Table~\ref{tab1} shows the parameters of the Gauss basis set, along with the basis spaces utilized in the coupled-cluster (CC) and configuration interaction (CI) calculations.
$N$ represents the number of the Gauss functions. $N_{\rm core}$ indicates the number of core orbitals. $N_{\rm valence}$ and $N_{\rm virtual}$ respectively denote the numbers of valence orbitals and virtual orbitals involved in the CC calculation. $N_{\rm CI}$ stands for the number of orbitals employed to build the configurations.
Specifically, single-particle orbitals with an energy lower than 20000 $a.u.$ are designated as virtual orbitals, and those with an energy lower than 500 $a.u.$ are used for constructing configurations. Nevertheless, in the MBPT calculations, the summation is conducted over the entire basis set. The sizes of the Gaussian basis set, virtual orbital space, and configurations quoted in Table~\ref{tab1} are already fully adequate: any further increase in basis set, virtual orbitals or configurations would shift the energies and hyperfine-structure parameters by less than one unit in the last reported digit. In short, within the present theoretical framework, the results are absolutely converged in basis set, virtual orbital, and configuration space.

To obtain the uncertainty of the results, we use five methods to construct the one-body and two-body correlation potentials. \textit{Method} 1 (CI+MBPT) derives these potentials via second-order many-ody perturbation theory. \textit{Method} 2 (CI+LCCSD) builds them through linear coupled-cluster singles and doubles calculations, and \textit{Method} 3 (CI+CCSD) determines them via full coupled-cluster singles and doubles calculations, with the latter considering non-linear terms related to single and double excitations of the cluster operator compared to CI+LCCSD. In \textit{Method} 2 and \textit{Method} 3, only contributions from single and double excited states are considered. To compensate for the overlooked higher-order correlation effects, a rescaling parameter $\rho_{\kappa}$ is introduced to replace the one-body correlation potential $\Sigma_{1}$ with $\rho_{\kappa}\Sigma_{1}$, and fine-tuning this parameter can bring the calculated energy closer to the experimental energy, a strategy previously used in CI+MBPT calculations ~\cite{Zhang2023PRA,Zhang2024PRA}. \textit{Method} 4 (CI+LCCSDs) starts from LCCSD-obtained potentials, keeps two-body potentials constant and applies a rescaling parameter to the one-body potential. \textit{Method} 5 (CI+CCSDs) does the same but starts from CCSD-obtained potentials. The rescaling parameters $\rho_{\kappa}$ for CI+LCCSDs and CI+CCSDs are listed in Table~\ref{Tab2}. Although CI+LCCSDs and CI+CCSDs methods depend on experimental energies, they remain valuable tools. They can provide high-lying excitation energies with appreciable accuracy, and by comparing results obtained with and without scaling parameters we can see how sensitive the property is to higher-order correlation, yielding a rough uncertainty estimate. Since the present work targets some low-lying states, we mainly use these two methods to assess the uncertainty of hyperfine-interaction parameters.
\begin{table}
\caption{The values of the rescaling parameters for CI+LCCSDs and CI+CCSDs methods.}\label{Tab2}
\begin{ruledtabular}
\begin{tabular}{ccccccccccccc}
\multicolumn{1}{c}{} & \multicolumn{2}{c}{Mg} & \multicolumn{2}{c}{Sr} & \multicolumn{2}{c}{Ba} \\
\cmidrule(lr){2-3} \cmidrule(lr){4-5} \cmidrule(lr){6-7}
$\rho_{\kappa}$ & LCCSDs & CCSDs & LCCSDs & CCSDs & LCCSDs & CCSDs \\
\hline
$\rho_{-1}$ & 1.011 & 1.035 & 0.938 & 0.988 & 0.924 & 0.985 \\
$\rho_{1}$ & 1.011 & 1.050 & 0.972 & 1.035 & 0.979 & 1.024 \\
$\rho_{-2}$ & 1.011 & 1.050 & 0.972 & 1.035 & 0.979 & 1.024 \\
$\rho_{2}$ & 1.00 & 1.000 & 0.957 & 1.045 & 0.946 & 1.041 \\
$\rho_{-3}$ & 1.00 & 1.000 & 0.957 & 1.045 & 0.946 & 1.041 \\
$\rho_{\rm others}$ & 1.00 & 1.000 & 1.000 & 1.000 & 1.000 & 1.000 \\
\end{tabular}
\end{ruledtabular}
\end{table}

\section{Results and Discussion}\label{results}

\subsection{Energy}

Table~\ref{Tab3} presents the energies (in cm$^{-1}$) of the low-lying states of Mg, Sr, and Ba atoms. These energies are obtained by using CI+MBPT(2), CI+LCCSD, CI+CCSD, CI+LCCSDs, and CI+CCSDs methods. To enhance the clarity and readability of the tabular data presentation, we have assigned the abbreviations $\rm M_{1}$, $\rm M_{2}$, $\rm M_{3}$, $\rm M_{4}$, and $\rm M_{5}$ to the CI+MBPT(2), CI+LCCSD, CI+CCSD, CI+LCCSDs, and CI+CCSDs methods, respectively.  We also compare our calculated results with the experimental values from National Institute of Standards and Technology (NIST)~\cite{NIST_ASD}. The symbol $\Delta_{n}$ represents the absolute difference between the theoretical results obtained by the $\rm M_{n}$ method and the experimental values, i.e., $\Delta_{n}=|E_{\rm M_{n}}-E_{\rm NIST}|$.

From Table~\ref{Tab3}, for the Mg atom, CI+MBPT, CI+LCCSD, and CI+CCSD all yield relatively accurate results, with differences from measured values less than 200 cm$^{-1}$. CI+LCCSD has the smallest difference, suggesting a cancellation between unconsidered correlation effects in CI+CCSD and non-linear terms of single and double excitations. As atomic number $Z$ increases for the three atoms, the difference between CI+MBPT results and measured values rises from 200 cm$^{-1}$ to 2500 cm$^{-1}$. The difference for CI+LCCSD results also increases but less significantly. CI+CCSD results are relatively stable, with differences within 250 cm$^{-1}$, possibly due to more complete correlation effects. CI+LCCSDs and CI+CCSDs, the two methods with scaling parameters, have even smaller differences (controlled within 100 cm$^{-1}$) from measured values.

\begin{small}
\begin{table*}
\caption[]{The energies (in cm$^{-1}$) of the low-lying states of Mg, Sr, and Ba atoms, calculated in CI+MBPT(2), CI+LCCSD, CI+CCSD, CI+LCCSDs, and CI+CCSDs approximation, are presented. The CI+MBPT(2), CI+LCCSD, CI+CCSD, CI+LCCSDs, and CI+CCSDs methods are denoted as $\rm M_{1}$, $\rm M_{2}$, $\rm M_{3}$, $\rm M_{4}$, and $\rm M_{5}$, respectively. The experimental values are taken from  NIST~\cite{NIST_ASD}. $\Delta_{n}$ represents the absolute difference between the theoretical results obtained by the $\rm M_{n}$ method and the experimental values, ie.
$\Delta_{n}=E_{\rm M_{n}}-E_{\rm NIST}$}\label{Tab3}
	\begin{ruledtabular}
		\begin{tabular}{cccccccccccccccc}
States& $E_{\rm M_{1}}$ & $E_{\rm M_{2}}$& $E_{\rm M_{3}}$& $E_{\rm M_{4}}$& $E_{\rm M_{5}}$& $E_{\rm NIST}$~\cite{NIST_ASD} & $\Delta_{1}$ & $\Delta_{2}$& $\Delta_{3}$& $\Delta_{4}$& $\Delta_{5}$ \\
\hline
\multicolumn{12}{c}{Mg}\\
3s3s  $^1\!S_0$      &  --182746&  --182903&  --182801&  --182942&  --182933&  --182939&  193&  36&  137&  3&  6 \\
3s3p  $^3\!P^o_0$ &  --160948&  --161051&  --160971&  --161081&  --161075&  --161088&  141&  37&  118&  7&  13 \\
3s3p  $^3\!P^o_1$ &  --160927&  --161032&  --160951&  --161062&  --161056&  --161068&  141&  36&  117&  6&  12 \\
3s3p  $^3\!P^o_2$ &  --160886   & --160990&  --160910&  --161020&  --161014&  --161027&141&  37&  118&  7&  13 \\
3s3p  $^1\!P^o_1$ &  --147712&  --147854&  --147766&  --147881&  --147861&  --147887&  176&  34&  122&  6&  26 \\
3s4s  $^3\!S_1$      &  --141600&  --141710&  --141635&  --141737&  --141727&  --141741&  141&  32&  106&  5&  15 \\
3s4s  $^1\!S_0$      &  --139294&  --139405&  --139331&  --139431&  --139422&  --139435&  141&  31&  104&  4&  13 \\
3s4p  $^3\!P^o_0$ &  --134965&  --135073&  --135001&  --135099&  --135090&  --135098&  133&  24&  96&  2&  8 \\
3s4p  $^3\!P^o_1$ &  --134962&  --135070&  --134998&  --135096&  --135086&  --135094&  133&  25&  96&  1&  8 \\
3s4p  $^3\!P^o_2$ &  --134955   & --135063&  --134991&  --135089&  --135080&  --135088&133&  24&  96&  2&  8 \\
3s4p  $^1\!P^o_1$ &  --133451&  --133569&  --133496&  --133594&  --133583&  --133592&  141&  23&  96&  2&  9 \\
\multicolumn{12}{c}{Sr}\\
5s5s  $^1\!S_0$      &  --136299&  --135478&  --135019&  --134932&  --134935&  --134897&  1401&  581&  121&  35&  38 \\
5s4d  $^3\!D_1$     &  --118249&  --117306&  --116517&  --116788&  --116712&  --116738&  1511&  568&  222&  49&  26 \\
5s4d  $^3\!D_2$    &  --118178&  --117239&  --116454&  --116723&  --116647&  --116679&  1500&  561&  225&  45&  32 \\
5s4d  $^3\!D_3$    &  --118060&  --117136&  --116355&  --116624&  --116544&  --116578&  1482&  558&  223&  45&  35 \\
5s4d  $^1\!D_2$    &  --116063&  --115211&  --114534&  --114748&  --114728&  --114748&  1315&  464&  213&  0&  20 \\
5s5p  $^3\!P^o_0$&  --121464&  --120904&  --120573&  --120549&  --120601&  --120580&  884&  324&  6&  31&  21 \\
5s5p  $^3\!P^o_1$&  --121278&  --120722&  --120393&  --120368&  --120419&  --120393&  885&  329&  0&  25&  26 \\
5s5p  $^3\!P^o_2$&  --120880&  --120321&  --119995&  --119970&  --120018&  --119999&  881&  322&  4&  29&  19 \\
5s5p  $^1\!P^o_1$&  --114345&  --113657&  --113237&  --113300&  --113271&  --113199&  1146&  458&  39&  102&  72 \\
\multicolumn{12}{c}{Ba} \\
6s6s  $^1\!S_0$     &  --125170&  --123485&  --122889&  --122753&  --122795&  --122721&  2449&  764&  168&  32&  74 \\
6s5d  $^3\!D_1$    &  --116127&  --114434&  --113473&  --113694&  --113689&  --113687&  2440&  747&  214&  7&  2 \\
6s5d  $^3\!D_2$    &  --115915&  --114242&  --113288&  --113507&  --113500&  --113505&  2409&  737&  218&  2&  6 \\
6s5d  $^3\!D_3$    &  --115464&  --113848&  --112906&  --113126&  --113107&  --113124&  2340&  723&  218&  1&  18 \\
6s5d  $^1\!D_2$    &  --113574&  --112012&  --111118&  --111324&  --111346&  --111326&  2248&  686&  207&  1&  20 \\
6s6p  $^3\!P^o_0$&  --111981&  --110881&  --110454&  --110428&  --110464&  --110455&  1526&  426&  1&  26&  9 \\
6s6p  $^3\!P^o_1$&  --111613&  --110516&  --110091&  --110065&  --110100&  --110084&  1529&  432&  6&  19&  15 \\
6s6p  $^3\!P^o_2$&  --110714&  --109627&  --109212&  --109180&  --109215&  --109206&  1508&  420&  6&  26&  9 \\
6s6p  $^1\!P^o_1$&  --106551&  --105211&  --104660&  --104754&  --104721&  --104661&  1890&  550&  1&  94&  61 \\
		\end{tabular}
	\end{ruledtabular}
\end{table*}
\end{small}
\subsection{Magnetic dipole hyperfine-structure constant }
Table~\ref{Tab4} showcases the hyperfine-structure constant $A$ (in MHz) for the $3s3p~^3\!P_{1,2}$ states of $^{25}$Mg, the $5s4d~^1\!D_{2}$ and $5s5p~^3\!P_{1,2}$ states of $^{87}$Sr,
 and the $6s5d~^3\!D_{1,2,3}$, $6s5d~^1\!D_{2}$, and $6s6p~^1\!P_{1}$ states of $^{137}$Ba. The data of nuclear spin $I$ and nuclear magnetic dipole moment $\mu$ are from
Ref.~\cite{Stone2005ADNDT}. Similar to Table~\ref{Tab3}, we also provide the calculation results of five methods. In the CI+MBPT calculation, the transition matrix elements include RPA and TP corrections. The remaining methods include RPA, HO, and TP corrections.
\begin{table}
	\caption[ ]{Hyperfine-structure constant $A$ (in MHz) of the low-lying states in $^{25}$Mg , $^{87}$Sr,
and $^{137}$Ba }\label{Tab4}
	\begin{ruledtabular}
		\begin{tabular}{clllllllllll}
States& $A_{\rm M_1}$  &$A_{\rm M_2}$&$A_{\rm M_3}$&$A_{\rm M_4}$&$A_{\rm M5}$&$A_{\rm Final}$\\
\hline
\multicolumn{7}{c}{$^{25}$Mg $(I=5/2,\mu=-0.85545\mu_{N})$ }\\
$3s4s$ $^3\!S_1$&--324 &--321 &--320 &--321 &--321 &--321(1)\\
$3s3p$ $^3\!P^o_1$&--146 &--143 &--143 &--143 &--143 &--143(1)\\
$3s3p$ $^3\!P^o_2$&--129 &--127 &--128 &--128 &--127 &--127(1)\\
\multicolumn{7}{c}{$^{87}$Sr $(I=9/2,\mu=-1.0936030\mu_{N})$}\\
$5s4d$ $^1\!D_2$&--0.99 &--5.28 &--5.51 &--5.44 &--5.36 &--5.3(0.3)\\
$5s5p$ $^3\!P^o_1$&--285 &--259 &--257 &--256 &--257 &--259(4)\\
$5s5p$ $^3\!P^o_2$&--235 &--213 &--211 &--210 &--210 &--213(4)\\
\multicolumn{7}{c}{$^{137}$Ba $(I=3/2,\mu=0.937365\mu_{N})$}\\
$6s5d$ $^3\!D_1$&--635 &--514 &--509 &--500 &--503 &--514(14)\\
$6s5d$ $^3\!D_2$&467 &421 &417 &410 &420 &421(11)\\
$6s5d$ $^1\!D_2$&--134 &--85 &--81 &--81 &--84 &--85 (4)\\
$6s5d$ $^3\!D_3$&507 &457 &455 &446 &452 &457(11)\\
$6s6p$ $^1\!P^o_1$&--151 &--112 &--110 &--111 &--111 &-112(2)\\
\end{tabular}
\end{ruledtabular}
\end{table}

From Table~\ref{Tab4}, for the $^{25}$Mg atom, the results of five methods are very close, especially the last four. As atomic number $Z$ increases, CI+MBPT results differ significantly from the other four. For instance, in the $5s4d$ $^1\!D_2$ state of $^{87}$Sr, CI+MBPT result is four-fold smaller, and the hfs constant $A$ of this state is two orders lower than those of other states. In the other seven states of $^{87}$Sr and $^{137}$Ba, discrepancies between CI+MBPT and the other four exceed $10\%$, reaching up to $50\%$ in the $5s4d$ $^1\!D_2$ state of $^{137}$Ba. We find that the results of CI+LCCSD, CI+CCSD, CI+LCCSDs, and CI+CCSDs are highly similar, with differences in $¡Ü 3\%$. As in our recent work ~\cite{Tang2025pra}, we use the CI+LCCSD result as the final recommended value. We compare it with the other three methods' results and take the maximum difference as the uncertainty. The final recommended value and its uncertainty are listed in the last column of Table~\ref{Tab4}.
\begin{table}
	\caption[ ]{Comparison between our results and available experimental and theoretical data of the hyperfine-structure constant
$A$.}\label{Tab5}
	\begin{ruledtabular}
		\begin{tabular}{ccccccccccccccccc}
States             &$A_{\rm Present}$     & $A_{\rm Expt.}$  & $A_{\rm Ther.}$\\
\hline
\multicolumn{4}{c}{$^{25}$Mg }\\
$3s4s$ $^3\!S_1$   &--321(1)    &--321.6(1.5)~\cite{HePRA2009}         &--325~\cite{HePRA2009}\\
$3s3p$ $^3\!P^o_1$ &--143(1)    &--144.945(5)~\cite{LurioPR1961}          &--141.365~\cite{Veseth1987JPhB},--146.1~\cite{Porsev2004PRA}\\
$3s3p$ $^3\!P^o_2$ &--127(1)    &--128.440(5)~\cite{LurioPR1961}          &--126.287~\cite{Veseth1987JPhB},--129.7~\cite{Porsev2004PRA}\\
                   &            &                                      &--127.5~\cite{Beloy2008PRA}\\
\multicolumn{4}{c}{$^{87}$Sr}\\
$5s4d$ $^1\!D_2$   &--5.3(0.3)  &--5.5734(4)~\cite{GrundevikZPA1983}   &\\
$5s5p$ $^3\!P^o_1$ &--259(4)    &--260.083(5)~\cite{PutlitzZP1963}     &--258.7~\cite{Porsev2004PRA},--256.96~\cite{Lu2019pra}\\
$5s5p$ $^3\!P^o_2$ &--213(4)    &--212.765(1)~\cite{HeiderPRA1977}     &--211.4~\cite{Porsev2004PRA},--230.6~\cite{Beloy2008PRA}\\
                   &            &                                      &--212.86~\cite{Lu2019pra}\\
\multicolumn{4}{c}{$^{137}$Ba}\\
$6s5d$ $^3\!D_1$   &--514(14)   &--520.602(6)~\cite{GustavssonZPA1979} &--547~\cite{Kozlov1999ejpd}\\
$6s5d$ $^3\!D_2$   &421(11)     &415.841(30)~\cite{GustavssonZPA1979}  &405~\cite{Kozlov1999ejpd}\\
$6s5d$ $^1\!D_2$   &--85 (4)    &--82.180(6)~\cite{SchmellingPRA1974}  &--102~\cite{Kozlov1999ejpd}\\
$6s5d$ $^3\!D_3$   &457(11)     &456.548(6)~\cite{GustavssonZPA1979}   &443~\cite{Kozlov1999ejpd}\\
$6s6p$ $^1\!P^o_1$ &-112(2)     &-109.50(13)~\cite{vanCJP1995}         &--107~\cite{Kozlov1999ejpd}\\
\end{tabular}
\end{ruledtabular}
\end{table}

Table~\ref{Tab5} compares our final recommended values of hyperfine-structure constants $A$ with available measured and theoretical results. We only list the most accurate measured values~\cite{LurioPR1961,HePRA2009,PutlitzZP1963,HeiderPRA1977,GrundevikZPA1983,GustavssonZPA1979,SchmellingPRA1974,vanCJP1995}. Considering uncertainty, our final recommended values match the measured ones. Except for the $5s4d$ $^1\!D_2$ state of $^{87}$Sr and the $5s5d$ $^1\!D_2$ state of $^{137}$Ba, the max difference between our recommended and measured results for other states is within 2\%, consistent with the $^{43}$Ca atom conclusion~\cite{Tang2025pra}. This further validates the reliability of the configuration interaction plus coupled- cluster method.

The last column of Table~\ref{Tab5} shows other theoretical results. For the $3s4s$ $^3\!S_1$ state of $^{25}$Mg, there's one available result from CI+MBPT calculation~\cite{HePRA2009}. Their method only considered RPA, while ours includes RPA, HO, and TP corrections. Considering only RPA, our CI+LCCSD result (323.8 MHz) differs from theirs (325 MHz) by about 1 MHz. For $3s3p$ $^3\!P^o_{1,2}$ states of $^{25}$Mg, Veseth's many-body perturbation theory results~\cite{Veseth1987JPhB} differ from ours by $1\%- 2\%$. The CI+MBPT results by Beloy \textit{et al.}~\cite{Beloy2008PRA} are consistent with ours, with a difference of less than $1$ MHz.  The CI+MBPT results by Porsev \textit{et al.}~\cite{Porsev2004PRA} also match our recommended values.

For the $^{87}$Sr atom, our final recommended values are very close to the MCDF results of Lu \textit{et al.}~\cite{Lu2019pra}, and the CI+MBPT results of Poserv \textit{et al.}~\cite{Porsev2004PRA}, with a difference of less than $1$ MHz. The CI+MBPT results of Beloy \textit{et al.}~\cite{Beloy2008PRA} are very close to our CI+MBPT results. We have not found that the hyperfine parameters of the $5s4d$ $^1\!D_2$ state have been calculated by other theoretical methods.
In contrast, in the $^{137}$Ba atom, the agreement between the CI+MBPT results of Kozlov \textit{et al.} ~\cite{Kozlov1999ejpd} and the experimental results is not as good as that of our CI+LCCSD results. This indicates that these states of the $^{137}$Ba atom are more sensitive to electron correlation effects, and it also shows that the electron correlation effects beyond the second-order many-body perturbation theory are important. In general, the configuration interaction plus coupled-cluster method can accurately calculate the hyperfine-structure properties of alkaline-earth atom systems.

\subsection{Nuclear quadrupole moment}

\begin{table}
	\caption[ ]{The electric-field gradients $q$ (in \textit{a.u.}) of the low-lying states in $^{25}$Mg , $^{87}$Sr,
and $^{137}$Ba. Multiply the electric-field gradient values by 234.9648867 to obtain MHz. }\label{Tab6}
	\begin{ruledtabular}
		\begin{tabular}{clllllllllll}
States  &$q_{\rm M_1}$&$q_{\rm M_2}$&$q_{\rm M_3}$&$q_{\rm M_4}$&$q_{\rm M_5}$&$q_{\rm Rec.}$\\
\hline
\multicolumn{7}{c}{Mg}\\
$3s3p$ $^3\!P^o_1$&--0.1749 &--0.1684 &--0.1671 &--0.1686 &--0.1677 &--0.1684(13)\\
$3s3p$ $^3\!P^o_2$&0.3482 &0.3352 &0.3326 &0.3355 &0.3337 &0.3352(26)\\
\multicolumn{7}{c}{Sr}\\
$5s4d$ $^1\!D_2$&0.7244 &0.6971 &0.7039 &0.6951 &0.7106 &0.697(14)\\
$5s5p$ $^3\!P^o_1$&--0.4939 &--0.4490 &--0.4415 &--0.4464 &--0.4452 &--0.4490(76)\\
$5s5p$ $^3\!P^o_2$&0.9359 &0.8525 &0.8381 &0.8476 &0.8451 &0.853(14)\\
\multicolumn{7}{c}{Ba}\\
$6s5d$ $^3\!D_1$&0.3042 &0.3100 &0.3161 &0.3066 &0.3194 &0.3100(94)\\
$6s5d$ $^3\!D_2$&0.4632 &0.4592 &0.4678 &0.4541 &0.4733 &0.459(14)\\
$6s5d$ $^1\!D_2$&1.0878 &1.0018 &1.0241 &0.9964 &1.0273 &1.002(26)\\
$6s5d$ $^3\!D_3$&0.8199 &0.8242 &0.8401 &0.8155 &0.8485 &0.824(24)\\
$6s6p$ $^1\!P^o_1$&1.0976 &0.8768 &0.8495 &0.8718 &0.8586 &0.877(27)\\
\end{tabular}
\end{ruledtabular}
\end{table}

Table~\ref{Tab6} shows electric-field gradients $q$ (in \textit{a.u.}) of low-lying states of Mg, Sr, and Ba atoms. To express the electric-field-gradient values in MHz, multiply them by 234.9648867. Like Table~\ref{Tab5}, it lists five methods' calculation results, final recommended values, and uncertainties. Like the hfs constant $A$, we adopt the CI+LCCSD result as our final recommended value. We compare it with the results from the other three methods and take the maximum difference as the uncertainty.
 For Mg, CI+MBPT method's results differ from the other four by about 3.8\%, larger than for hfs constants $A$; differences among the other four are within 1\%. In Sr's $5s5p$ $^3P^o_1$ and $5s5p$ $^3P^o_2$ states, CI+MBPT results differ from the other four by about 10\%, while the other four differ within 2\%, similar to the hfs constant $A$ results in Table~\ref{Tab4}. For Sr's $5s4d$ $^1\!D_2$ state, the electric-field gradient's electron-correlation trend differs from the hfs constant $A$. CI+MBPT results differ from the other four by about 4\%, but for hfs constant $A$, CI+MBPT result is four times smaller. In Ba's $6s5d$ $^3\!D_1$, $6s5d$ $^3\!D_2$, and $6s5d$ $^3\!D_3$ states, the five methods' results differ within 3\%, and CI+MBPT result differs from the final recommended value within 2\%. For Ba's $6s5d$ $^1\!D_2$ and $6s6p$ $^1\!P^o_1$ singlet states, CI+MBPT results differ from the final recommended values by 8.5\% and 25\% respectively, showing importance of correlation effects beyond MBPT(2). The other four methods' results differ within 3\%. Comparing Table~\ref{Tab4} and Table~\ref{Tab6}, there are differences in electron correlation trends between magnetic-dipole and electric-quadrupole hyperfine interactions.

\begin{table}
	\caption[ ]{The nuclear quadrupole moment $Q$ (in b) of $^{25}$Mg , $^{87}$Sr, $^{135}$Ba and $^{137}$Ba.}\label{Tab7}
	\begin{ruledtabular}
		\begin{tabular}{ccccccccc}
State            &$B_{\rm Expt.}$ & $Q$ \\
\hline
\multicolumn{3}{c}{$^{25}$Mg}\\
$3s3p$ $^3\!P^o_1$&--8.029(5)~\cite{LurioPR1961}&0.2029(21)\\
$3s3p$ $^3\!P^o_2$&16.009(5)~\cite{LurioPR1961}&0.2032(18)\\
Final result          &      &0.203(2) \\
\multicolumn{3}{c}{$^{87}$Sr}\\
$5s4d$ $^1\!D_2$   &55.421(6)~\cite{GrundevikZPA1983}&0.3384(62)\\
$5s5p$ $^3\!P^o_1$ &--35.355(6)~\cite{PutlitzZP1963}&0.3351(64)\\
$5s5p$ $^3\!P^o_2$ &67.215(15)~\cite{HeiderPRA1977}&0.3356(67)\\
Final result           &                               &0.336(4)\\
\multicolumn{3}{c}{$^{137}$Ba}\\
$6s5d$ $^3\!D_1$&17.950(8)~\cite{GustavssonZPA1979}&0.2456(74)\\
$6s5d$ $^3\!D_2$&25.756(50)~\cite{GustavssonZPA1979}&0.2400(89)\\
$6s5d$ $^1\!D_2$&59.568(19)~\cite{SchmellingPRA1974}&0.2531(65)\\
$6s5d$ $^3\!D_3$&47.302(60)~\cite{GustavssonZPA1979}&0.2447(76)\\
$6s6p$ $^1\!P^o_1$&50.09(21)~\cite{vanCJP1995}&0.2431(83)\\
Final result             &                           &0.246(4) \\
\multicolumn{3}{c}{$^{135}$Ba} \\
$6s5d$ $^3\!D_1$&11.690(8)~\cite{GustavssonZPA1979}&0.1605(48)\\
$6s5d$ $^3\!D_2$&16.522(54)~\cite{GustavssonZPA1979}&0.1531(57)\\
$6s5d$ $^1\!D_2$&38.713(20)~\cite{SchmellingPRA1974}&0.1645(42)\\
$6s5d$ $^3\!D_3$&30.800(62)~\cite{GustavssonZPA1979}&0.1590(49)\\
$6s6p$ $^1\!P^o_1$&34.01(22)~\cite{vanCJP1995}&0.1651(56)\\
Final result          &                           &0.161(2)\\
\end{tabular}
\end{ruledtabular}
\end{table}

By combining the electric-field gradients in Table~\ref{Tab6} and the measured electric quadrupole hyperfine-structure constants of these low-lying states~\cite{LurioPR1961,GrundevikZPA1983,PutlitzZP1963,HeiderPRA1977,GustavssonZPA1979,SchmellingPRA1974,vanCJP1995}, we obtain the nuclear quadrupole moments of $^{25}$Mg, $^{87}$Sr, $^{135}$Ba and $^{137}$Ba, which are listed in Table~\ref{Tab7}. In Table~\ref{Tab7}, we not only present the nuclear quadrupole moments extracted from individual states, but also provide the final recommended values and their uncertainties. The final recommended values and their uncertainties are obtained by the following equations:
\begin{align}
\begin{cases}
Q_{\rm Final}=\frac{\sum_{i}{\frac{Q_{i}}{(\Delta{Q_{i}})^2}}}{\sum_{i}{\frac{1}{(\Delta{Q_{i}})^2}}}\\
\Delta{Q_{\rm Final}}=\frac{1}{\sqrt{\sum_{i}{\frac{1}{(\Delta{Q_{i}})^2}}}}
\end{cases},
\end{align}
where $Q_{i}$ and $\Delta{Q_{i}}$ denote the nuclear quadrupole moment and its uncertainty obtained from the $i$-th state, respectively. The accuracy of these measured hfs constants $B$ is high. Therefore, the uncertainties of the nuclear quadrupole moments we extracted are entirely due to the electric-field gradients.

For $^{25}$Mg, the nuclear quadrupole moments extracted from the $3s3p$ $^3\!P^o_1$ and $3s3p$ $^3\!P^o_2$ states are very close to the final recommended value. The differences among the three are less than 0.2\%, and the uncertainty of the final recommended value is approximately 1\%. For $^{87}$Sr, the differences among the nuclear quadrupole moments obtained from $5s4d~^1\!D_{2}$ and $5s5p~^3\!P_{1,2}$ states are within 1\%.

 Compared with the Sr atom, the differences among the nuclear quadrupole moments of $^{137}$Ba and $^{135}$Ba extracted from the $6s5d$ $^3\!D_1$, $6s5d$ $^3\!D_2$, and $6s5d$ $^3\!D_3$ states are more obvious. For $^{137}$Ba, the nuclear quadrupole moment extracted from the $6s5d$ $^1\!D_2$ state is significantly larger than those from other states. The difference between the nuclear quadrupole moments extracted from single states and the final recommended values is within 3\%. However, for $^{135}$Ba, the nuclear quadrupole moment extracted from the $6s5d$ $^3\!D_2$ state is significantly smaller than those from the other four states, and is smaller than the final recommended value by about 5\%. We observe that in the $6s5d$ $^3\!D_1$, $6s5d$ $^1\!D_2$, and $6s5d$ $^3\!D_3$ states, the ratios of the hfs constants $B$ of $^{137}$Ba to $^{135}$Ba are $1.5355$, $1.5387$, and $1.5358$ respectively, which are consistent with the measured value of 1.538485(95) by the nuclear-magnetic-resonance techniques~\cite{Lutz1978ZPA}. However, the ratio for the $6s5d$ $^3\!D_2$ state is 1.5589, which is 1.3\% larger than the measured value~\cite{Lutz1978ZPA}. In the $6s6p$ $^1\!P^o_1$ state, the ratio of the hfs constants $B$ of $^{137}$Ba to $^{135}$Ba is $1.4728$, deviating from the measured value~\cite{Lutz1978ZPA} by approximately 4\%. Compared with the other four states, the measurement precision of the $6s6p$ $^1\!P^o_1$ state is nearly one order of magnitude lower. The ratio of the nuclear quadrupole moments of $^{137}$Ba to $^{135}$Ba we obtained through our final recommended values is approximately 1.530, which is smaller than the measured value~\cite{Lutz1978ZPA} by about 0.5\%.

\begin{table}
	\caption[ ]{Comparison between our calculated values and available experimental and theoretical data of the nuclear quadrupole moment $Q$ (in b) of $^{25}$Mg , $^{87}$Sr, and $^{137}$Ba.}\label{Tab8}
	\begin{ruledtabular}
		\begin{tabular}{clllll}
$Q$ &\multicolumn{1}{c}{Sources} \\
\hline
\multicolumn{2}{c}{$^{25}$Mg}\\
0.203(2)     &Present, $3s3p$ $^3\!P^o_1$, $3s3p$ $^3\!P^o_2$\\
0.1994(20)   &~\cite{Sundholm1991npa}, $3s3p$ $^3\!P^o_2$\\
0.201(3)     &~\cite{Weber1982npa}, Muonic X-ray experiment\\
0.196(4)     &~\cite{Schwalm1977npa}, Nuclear scattering\\
\\
\multicolumn{2}{c}{$^{87}$Sr}\\
0.336(4)     &Present, $5s4d$ $^1\!D_2$, $5s5p$ $^3\!P^o_{1}$, $5s5p$ $^3\!P^o_{2}$\\
0.305(2)     &~\cite{Sahoo2006pra}, $4d_{5/2}$ \\
0.328(4)     &~\cite{Lu2019pra}, $5s5p$ $^3\!P^o_{1}$, $5s5p$ $^3\!P^o_{2}$   \\
0.335(20)    &~\cite{HeiderPRA1977}, $5s5p$ $^3\!P^o_{1}$, $5s5p$ $^3\!P^o_{2}$  \\
0.327(24)    &~\cite{Pendrill2002JPhB}, $5p_{3/2}$ \\
0.323(20)    &~\cite{Yu2004PRA}, $5p_{3/2}$ \\
0.3102(19)   &~\cite{Chakraborty2025arxiv}, $4d_{5/2}$ \\
\\
\multicolumn{2}{c}{$^{137}$Ba}\\
0.246(4)   &Present, $6s5d$ $^3\!D_{1,2,3}$, $6s5d$ $^1\!D_2$, $6s6p$ $^1\!P^o_1$ \\
0.236(3)   &~\cite{Sahoo2013pra},  $5d_{3/2}$, $5d_{5/2}$ \\
0.246(2)   &~\cite{Silverans1986pra}, $6p_{3/2}$ \\
0.245(4)   &~\cite{Wendt1988ZPA}, $6p_{3/2}$ \\
0.246(1)   &~\cite{Sahoo2006PRAa}, $6p_{3/2}$ \\
0.2344(17) &~\cite{Chakraborty2025arxiv},  $5d_{3/2}$,$5d_{5/2}$ \\
\end{tabular}
\end{ruledtabular}
\end{table}

Table~\ref{Tab8} compares our nuclear quadrupole moments with other experimental and theoretical results~\cite{Sundholm1991npa,Weber1982npa,Schwalm1977npa,Sahoo2006pra,Lu2019pra,HeiderPRA1977,Pendrill2002JPhB,
Yu2004PRA,Chakraborty2025arxiv,Sahoo2013pra,Silverans1986pra,Wendt1988ZPA}. For $^{25}$Mg, considering uncertainties, our results agree well with those from the $3s3p$ $^3\!P^o_2$ state via finite-element multi-configuration Dirac-Fock calculation~\cite{Sundholm1991npa} and muonic X-ray experiment~\cite{Weber1982npa}, while the result from nuclear scattering~\cite{Schwalm1977npa} is slightly smaller.
For $^{87}$Sr, we list the nuclear quadrupole moments extracted from the $5p_{3/2}$ and $4d_{5/2}$ states of $^{87}$Sr$^{+}$~\cite{Pendrill2002JPhB,Yu2004PRA,Sahoo2006pra,Chakraborty2025arxiv}, and those extracted from the $5s5p$ $^3\!P^o_{1}$ and $5s5p$ $^3\!P^o_{2}$ states of neutral $^{87}$Sr~\cite{HeiderPRA1977,Lu2019pra}. Our results are consistent with the nuclear quadrupole moments extracted from the $5p_{3/2}$ state~\cite{Pendrill2002JPhB,Yu2004PRA}, as well as those from the $5s5p$ $^3\!P^o_{1}$ and $5s5p$ $^3\!P^o_{2}$ states~\cite{HeiderPRA1977,Lu2019pra}. Notably, the uncertainty of our results is smaller than that of the nuclear quadrupole moment determined from the $5p_{3/2}$ state~\cite{Pendrill2002JPhB,Yu2004PRA}, and is comparable to the uncertainty evaluated by Lu \textit{et al.} using the MCDF method~\cite{Lu2019pra}. However, all these results are larger than the nuclear quadrupole moment extracted from the $4d_{5/2}$ state~\cite{Sahoo2013pra}, and this difference is similar to what we previously observed in $^{43}$Ca$^{+}$ and $^{43}$Ca~\cite{Tang2025pra}. Recently, Chakraborty \textit{et al.} recalculated the electric-field gradient of the $4d_{5/2}$ state using the coupled-cluster singles, doubles, and triples method and redetermined the nuclear quadrupole moment of $^{87}$Sr precisely~\cite{Chakraborty2025arxiv}. Their focus on triple excitations leads to high-accuracy electric-field gradients, so the measured hyperfine-structure constants of the $^{87}$Sr$^{+}$ $4d_{5/2}$ state need further confirmation.

For $^{137}$Ba, we compare our results with the nuclear quadrupole moments extracted from the $6p_{3/2}$ and $5d_{3/2,5/2}$ states of $^{137}$Ba$^{+}$~\cite{Silverans1986pra, Wendt1988ZPA,Sahoo2006PRAa,Sahoo2013pra, Chakraborty2025arxiv}. In this table, we do not list the nuclear quadrupole moments extracted from the $6p5d$ and $5s5p$ configurations of neutral Ba atoms in earlier works~\cite{Jackson964rsa,GustavssonZPA1979,Grundevik1982zpa}. This is because the theoretical values of the hyperfine-structure parameters at that time had low precision, resulting in imprecise nuclear quadrupole moments or large uncertainties.

The comparison shows that our result is in excellent agreement with the nuclear quadrupole moments extracted from the $6p_{3/2}$ state of $^{137}$Ba$^{+}$~\cite{Silverans1986pra, Wendt1988ZPA,Sahoo2006PRAa}. However, there is a deviation of about 4\% compared with the results extracted from the $5d_{3/2,5/2}$ states of $^{137}$Ba$^{+}$~\cite{Sahoo2013pra, Chakraborty2025arxiv}. For the $6p_{3/2}$ state, two groups independently carried out the measurement of its electric quadrupole hyperfine-structure constants~\cite{Wendt1988ZPA,Villemoes1993jpb}. The measurement results they obtained are in very good agreement, and the measurement precision of both groups is better than 1\%. Similarly, for the $5d_{3/2,5/2}$ states, two teams independently measured the hyperfine splittings of these two states~\cite{Silverans1986pra,Lewty_2013_pra}. The values of the hfs constants $B$ they obtained are highly consistent, and the measurement precision of the $5d_{3/2,5/2}$ states is at least an order of magnitude higher than that of the $6p_{3/2}$ state. If the theoretical framework can accurately provide the electric-field gradient, then by combining the hfs constants $B$ of the $6p_{3/2}$ or $5d_{3/2,5/2}$ states, the nuclear quadrupole moment can be extracted accurately. In the recent work by Chakraborty \textit{et al.}~\cite{Chakraborty2025arxiv}, they used the coupled-cluster singles, doubles, and triples method to calculate the electric-field gradient of the $5d_{3/2,5/2}$ states of $^{137}$Ba$^{+}$, and obtained an accurate nuclear quadrupole moment of $^{137}$Ba. The uncertainty of their result is extremely small. Our result is obtained by statistically averaging the results of five states of the $^{137}$Ba atom. Except for the  $6s5d$ $^1\!D_2$ state, the difference between our calculated magnetic-dipole hyperfine-structure constant and the measured value is within 2\% as shown in Table~\ref{Tab5}. Based on this, we believe that our result should be reliable. However, there is still room for improvement in our method. For example, the contribution of the triple-excitation is not currently included in our CC calculation, and the three-body interaction potential is also ignored, which may be one of the reasons for the 4\% difference. The recent work by Chakraborty \textit{et al.} shows that triple excitations account for $1-2\%$ of the hfs constants $B$ of the $5d_{3/2}$ and $5d_{5/2}$ stetes in Ba$^{+}$~\cite{Chakraborty2025arxiv}. A contribution of this size is expected to be equally relevant for neutral alkaline-earth atoms. Since the situation of
$^{135}$Ba is the same as that of $^{137}$Ba, we only compared $^{137}$Ba in Table~\ref{Tab8}.

\section{Conclusion}\label{conclusions}
In this work, we calculated the magnetic dipole hyperfine-structure constants and electric-field gradients for the $3s3p~^3\!P_{1,2}$ states of $^{24}$Mg, the $5s4d~^1\!D_{2}$ and $5s5p~^3\!P_{1,2}$ states of $^{87}$Sr, and the $6s5d~^3\!D_{1,2,3}$, $6s5d~^1\!D_{2}$, and $6s6p~^1\!P_{1}$ states of $^{135,137}$Ba. We utilized five computational methods, namely CI+MBPT, CI+LCCSD, CI+CCSD, CI+LCCSDs, and CI+CCSDs, to derive the final recommended values along with their corresponding uncertainties. Our final recommended values for the magnetic dipole hyperfine-structure constants of these states are fully consistent with the measured results considering uncertainties.

By combining the measured electric-quadrupole hyperfine-structure constants with our calculated electric-field gradients for the relevant states, we obtain nuclear quadrupole moments of 0.203(2) b, 0.336(4) b, 0.161(2) b and 0.246(4) b for $^{25}$Mg, $^{87}$Sr, $^{135}$Ba, and $^{137}$Ba, respectively. Our value for $^{25}$Mg agrees with the muonic X-ray result considering the uncertainty. The $^{87}$Sr quadrupole moment is consistent with earlier extractions from the $5p_{3/2}$ state of $^{87}$Sr$^{+}$ as well as from the  the $5s5p$ $^3\!P^o_{1}$ and $5s5p$ $^3\!P^o_{2}$ states of $^{87}$Sr, and the nuclear quadrupole moments of $^{135,137}$Ba agree with those derived from the $6p_{3/2}$ state of $^{135,137}$Ba$^{+}$. Nevertheless, there are approximately 10\% and 4\% discrepancies between our nuclear quadrupole moments of $^{87}$Sr and $^{137}$Ba and the currently adopted values~\cite{Pekka2018MP} obtained from the $4d_{5/2}$ state of $^{87}$Sr$^{+}$ and the $5d_{3/2,5/2}$ state of $^{137}$Ba$^{+}$. These discrepancies between neutral-atom- and ion-based extractions remain to be clarified by further experimental and theoretical works. The recent work indicates that the triple excitations contribute $1-2\%$ to the electric-quadrupole hyperfine-structure constants of the $d_{3/2}$ and $d_{5/2}$ states in Sr$^{+}$ and Ba$^{+}$~\cite{Chakraborty2025arxiv}. To enhance further the accuracy of nuclear quadrupole moments derived with the RCI+CC calculations, the triple excitation contributions need to be included.

\begin{acknowledgments}
The work was supported by the National Natural Science Foundation of China under Grant No.12174268.
\end{acknowledgments}


\begin{thebibliography}{60}%
\makeatletter
\providecommand \@ifxundefined [1]{%
 \@ifx{#1\undefined}
}%
\providecommand \@ifnum [1]{%
 \ifnum #1\expandafter \@firstoftwo
 \else \expandafter \@secondoftwo
 \fi
}%
\providecommand \@ifx [1]{%
 \ifx #1\expandafter \@firstoftwo
 \else \expandafter \@secondoftwo
 \fi
}%
\providecommand \natexlab [1]{#1}%
\providecommand \enquote  [1]{``#1''}%
\providecommand \bibnamefont  [1]{#1}%
\providecommand \bibfnamefont [1]{#1}%
\providecommand \citenamefont [1]{#1}%
\providecommand \href@noop [0]{\@secondoftwo}%
\providecommand \href [0]{\begingroup \@sanitize@url \@href}%
\providecommand \@href[1]{\@@startlink{#1}\@@href}%
\providecommand \@@href[1]{\endgroup#1\@@endlink}%
\providecommand \@sanitize@url [0]{\catcode `\\12\catcode `\$12\catcode
  `\&12\catcode `\#12\catcode `\^12\catcode `\_12\catcode `\%12\relax}%
\providecommand \@@startlink[1]{}%
\providecommand \@@endlink[0]{}%
\providecommand \url  [0]{\begingroup\@sanitize@url \@url }%
\providecommand \@url [1]{\endgroup\@href {#1}{\urlprefix }}%
\providecommand \urlprefix  [0]{URL }%
\providecommand \Eprint [0]{\href }%
\providecommand \doibase [0]{http://dx.doi.org/}%
\providecommand \selectlanguage [0]{\@gobble}%
\providecommand \bibinfo  [0]{\@secondoftwo}%
\providecommand \bibfield  [0]{\@secondoftwo}%
\providecommand \translation [1]{[#1]}%
\providecommand \BibitemOpen [0]{}%
\providecommand \bibitemStop [0]{}%
\providecommand \bibitemNoStop [0]{.\EOS\space}%
\providecommand \EOS [0]{\spacefactor3000\relax}%
\providecommand \BibitemShut  [1]{\csname bibitem#1\endcsname}%
\let\auto@bib@innerbib\@empty
\bibitem [{\citenamefont {Qi}\ \emph {et~al.}(2023)\citenamefont {Qi},
  \citenamefont {Zhang}, \citenamefont {Yan}, \citenamefont {Shi},
  \citenamefont {Drake}, \citenamefont {Chen},\ and\ \citenamefont
  {Zhong}}]{QiPRA2023}%
  \BibitemOpen
  \bibfield  {author} {\bibinfo {author} {\bibfnamefont {X.-Q.}\ \bibnamefont
  {Qi}}, \bibinfo {author} {\bibfnamefont {P.-P.}\ \bibnamefont {Zhang}},
  \bibinfo {author} {\bibfnamefont {Z.-C.}\ \bibnamefont {Yan}}, \bibinfo
  {author} {\bibfnamefont {T.-Y.}\ \bibnamefont {Shi}}, \bibinfo {author}
  {\bibfnamefont {G.~W.~F.}\ \bibnamefont {Drake}}, \bibinfo {author}
  {\bibfnamefont {A.-X.}\ \bibnamefont {Chen}}, \ and\ \bibinfo {author}
  {\bibfnamefont {Z.-X.}\ \bibnamefont {Zhong}},\ }\href {\doibase
  10.1103/PhysRevA.107.L010802} {\bibfield  {journal} {\bibinfo  {journal}
  {Phys. Rev. A}\ }\textbf {\bibinfo {volume} {107}},\ \bibinfo {pages}
  {L010802} (\bibinfo {year} {2023})}\BibitemShut {NoStop}%
\bibitem [{\citenamefont {Qi}\ \emph {et~al.}(2025)\citenamefont {Qi},
  \citenamefont {Zhang}, \citenamefont {Yan}, \citenamefont {Tang},
  \citenamefont {Chen}, \citenamefont {Shi},\ and\ \citenamefont
  {Zhong}}]{QiPRR2025}%
  \BibitemOpen
  \bibfield  {author} {\bibinfo {author} {\bibfnamefont {X.-Q.}\ \bibnamefont
  {Qi}}, \bibinfo {author} {\bibfnamefont {P.-P.}\ \bibnamefont {Zhang}},
  \bibinfo {author} {\bibfnamefont {Z.-C.}\ \bibnamefont {Yan}}, \bibinfo
  {author} {\bibfnamefont {L.-Y.}\ \bibnamefont {Tang}}, \bibinfo {author}
  {\bibfnamefont {A.-X.}\ \bibnamefont {Chen}}, \bibinfo {author}
  {\bibfnamefont {T.-Y.}\ \bibnamefont {Shi}}, \ and\ \bibinfo {author}
  {\bibfnamefont {Z.-X.}\ \bibnamefont {Zhong}},\ }\href {\doibase
  10.1103/PhysRevResearch.7.L022020} {\bibfield  {journal} {\bibinfo  {journal}
  {Phys. Rev. Res.}\ }\textbf {\bibinfo {volume} {7}},\ \bibinfo {pages}
  {L022020} (\bibinfo {year} {2025})}\BibitemShut {NoStop}%
\bibitem [{\citenamefont {Guan}\ \emph {et~al.}(2025)\citenamefont {Guan},
  \citenamefont {Qi}, \citenamefont {Li}, \citenamefont {Zhou}, \citenamefont
  {Sun}, \citenamefont {Chen}, \citenamefont {Chang}, \citenamefont {Huang},
  \citenamefont {Zhang}, \citenamefont {Yan}, \citenamefont {Drake},
  \citenamefont {Chen}, \citenamefont {Zhong}, \citenamefont {Wang},
  \citenamefont {Michel}, \citenamefont {Shi},\ and\ \citenamefont
  {Gao}}]{GuanPRAL2025}%
  \BibitemOpen
  \bibfield  {author} {\bibinfo {author} {\bibfnamefont {H.}~\bibnamefont
  {Guan}}, \bibinfo {author} {\bibfnamefont {X.-Q.}\ \bibnamefont {Qi}},
  \bibinfo {author} {\bibfnamefont {J.-G.}\ \bibnamefont {Li}}, \bibinfo
  {author} {\bibfnamefont {P.-P.}\ \bibnamefont {Zhou}}, \bibinfo {author}
  {\bibfnamefont {W.}~\bibnamefont {Sun}}, \bibinfo {author} {\bibfnamefont
  {S.-L.}\ \bibnamefont {Chen}}, \bibinfo {author} {\bibfnamefont {X.-R.}\
  \bibnamefont {Chang}}, \bibinfo {author} {\bibfnamefont {Y.}~\bibnamefont
  {Huang}}, \bibinfo {author} {\bibfnamefont {P.-P.}\ \bibnamefont {Zhang}},
  \bibinfo {author} {\bibfnamefont {Z.-C.}\ \bibnamefont {Yan}}, \bibinfo
  {author} {\bibfnamefont {G.~W.~F.}\ \bibnamefont {Drake}}, \bibinfo {author}
  {\bibfnamefont {A.-X.}\ \bibnamefont {Chen}}, \bibinfo {author}
  {\bibfnamefont {Z.-X.}\ \bibnamefont {Zhong}}, \bibinfo {author}
  {\bibfnamefont {J.-L.}\ \bibnamefont {Wang}}, \bibinfo {author}
  {\bibfnamefont {N.}~\bibnamefont {Michel}}, \bibinfo {author} {\bibfnamefont
  {T.-Y.}\ \bibnamefont {Shi}}, \ and\ \bibinfo {author} {\bibfnamefont
  {K.-L.}\ \bibnamefont {Gao}},\ }\href {\doibase 10.1103/s61z-b1t5} {\bibfield
   {journal} {\bibinfo  {journal} {Phys. Rev. A}\ }\textbf {\bibinfo {volume}
  {112}},\ \bibinfo {pages} {L010801} (\bibinfo {year} {2025})}\BibitemShut
  {NoStop}%
\bibitem [{\citenamefont {{Heyde}}\ and\ \citenamefont
  {{Wood}}(2011)}]{Heyde2011RMP}%
  \BibitemOpen
  \bibfield  {author} {\bibinfo {author} {\bibfnamefont {K.}~\bibnamefont
  {{Heyde}}}\ and\ \bibinfo {author} {\bibfnamefont {J.~L.}\ \bibnamefont
  {{Wood}}},\ }\href {\doibase 10.1103/RevModPhys.83.1467} {\bibfield
  {journal} {\bibinfo  {journal} {Rev. Mod. Phys.}\ }\textbf {\bibinfo {volume}
  {83}},\ \bibinfo {pages} {1467} (\bibinfo {year} {2011})}\BibitemShut
  {NoStop}%
\bibitem [{\citenamefont {{Campbell}}\ \emph {et~al.}(2016)\citenamefont
  {{Campbell}}, \citenamefont {{Moore}},\ and\ \citenamefont
  {{Pearson}}}]{Campbell2016PRNP}%
  \BibitemOpen
  \bibfield  {author} {\bibinfo {author} {\bibfnamefont {P.}~\bibnamefont
  {{Campbell}}}, \bibinfo {author} {\bibfnamefont {I.~D.}\ \bibnamefont
  {{Moore}}}, \ and\ \bibinfo {author} {\bibfnamefont {M.~R.}\ \bibnamefont
  {{Pearson}}},\ }\href {\doibase 10.1016/j.ppnp.2015.09.003} {\bibfield
  {journal} {\bibinfo  {journal} {Prog. Part. Nucl. Phys.}\ }\textbf {\bibinfo
  {volume} {86}},\ \bibinfo {pages} {127} (\bibinfo {year} {2016})}\BibitemShut
  {NoStop}%
\bibitem [{\citenamefont {Sinitsyn}\ \emph {et~al.}(2012)\citenamefont
  {Sinitsyn}, \citenamefont {Li}, \citenamefont {Crooker}, \citenamefont
  {Saxena},\ and\ \citenamefont {Smith}}]{Sinitsyn2012PRL}%
  \BibitemOpen
  \bibfield  {author} {\bibinfo {author} {\bibfnamefont {N.~A.}\ \bibnamefont
  {Sinitsyn}}, \bibinfo {author} {\bibfnamefont {Y.}~\bibnamefont {Li}},
  \bibinfo {author} {\bibfnamefont {S.~A.}\ \bibnamefont {Crooker}}, \bibinfo
  {author} {\bibfnamefont {A.}~\bibnamefont {Saxena}}, \ and\ \bibinfo {author}
  {\bibfnamefont {D.~L.}\ \bibnamefont {Smith}},\ }\href {\doibase
  10.1103/PhysRevLett.109.166605} {\bibfield  {journal} {\bibinfo  {journal}
  {Phys. Rev. Lett.}\ }\textbf {\bibinfo {volume} {109}},\ \bibinfo {pages}
  {166605} (\bibinfo {year} {2012})}\BibitemShut {NoStop}%
\bibitem [{\citenamefont {{Pyykk{\"o}}}(2018)}]{Pekka2018MP}%
  \BibitemOpen
  \bibfield  {author} {\bibinfo {author} {\bibfnamefont {P.}~\bibnamefont
  {{Pyykk{\"o}}}},\ }\href {\doibase 10.1080/00268976.2018.1426131} {\bibfield
  {journal} {\bibinfo  {journal} {Mol. Phys.}\ }\textbf {\bibinfo {volume}
  {116}},\ \bibinfo {pages} {1328} (\bibinfo {year} {2018})}\BibitemShut
  {NoStop}%
\bibitem [{\citenamefont {{Wong}}\ \emph {et~al.}(2006)\citenamefont {{Wong}},
  \citenamefont {{Howes}}, \citenamefont {{Dupree}},\ and\ \citenamefont
  {{Smith}}}]{WongCPL2006}%
  \BibitemOpen
  \bibfield  {author} {\bibinfo {author} {\bibfnamefont {A.}~\bibnamefont
  {{Wong}}}, \bibinfo {author} {\bibfnamefont {A.~P.}\ \bibnamefont {{Howes}}},
  \bibinfo {author} {\bibfnamefont {R.}~\bibnamefont {{Dupree}}}, \ and\
  \bibinfo {author} {\bibfnamefont {M.~E.}\ \bibnamefont {{Smith}}},\ }\href
  {\doibase 10.1016/j.cplett.2006.06.039} {\bibfield  {journal} {\bibinfo
  {journal} {Chem. Phys. Lett.}\ }\textbf {\bibinfo {volume} {427}},\ \bibinfo
  {pages} {201} (\bibinfo {year} {2006})}\BibitemShut {NoStop}%
\bibitem [{\citenamefont {Sahoo}\ \emph {et~al.}(2013)\citenamefont {Sahoo},
  \citenamefont {Barrett},\ and\ \citenamefont {Das}}]{Sahoo2013pra}%
  \BibitemOpen
  \bibfield  {author} {\bibinfo {author} {\bibfnamefont {B.~K.}\ \bibnamefont
  {Sahoo}}, \bibinfo {author} {\bibfnamefont {M.~D.}\ \bibnamefont {Barrett}},
  \ and\ \bibinfo {author} {\bibfnamefont {B.~P.}\ \bibnamefont {Das}},\ }\href
  {\doibase 10.1103/PhysRevA.87.042506} {\bibfield  {journal} {\bibinfo
  {journal} {Phys. Rev. A}\ }\textbf {\bibinfo {volume} {87}},\ \bibinfo
  {pages} {042506} (\bibinfo {year} {2013})}\BibitemShut {NoStop}%
\bibitem [{\citenamefont {{Biero{\'n}}}\ \emph {et~al.}(2018)\citenamefont
  {{Biero{\'n}}}, \citenamefont {{Filippin}}, \citenamefont {{Gaigalas}},
  \citenamefont {{Godefroid}}, \citenamefont {{J{\"o}nsson}},\ and\
  \citenamefont {{Pyykk{\"o}}}}]{Bi2018pra}%
  \BibitemOpen
  \bibfield  {author} {\bibinfo {author} {\bibfnamefont {J.}~\bibnamefont
  {{Biero{\'n}}}}, \bibinfo {author} {\bibfnamefont {L.}~\bibnamefont
  {{Filippin}}}, \bibinfo {author} {\bibfnamefont {G.}~\bibnamefont
  {{Gaigalas}}}, \bibinfo {author} {\bibfnamefont {M.}~\bibnamefont
  {{Godefroid}}}, \bibinfo {author} {\bibfnamefont {P.}~\bibnamefont
  {{J{\"o}nsson}}}, \ and\ \bibinfo {author} {\bibfnamefont {P.}~\bibnamefont
  {{Pyykk{\"o}}}},\ }\href {\doibase 10.1103/PhysRevA.97.062505} {\bibfield
  {journal} {\bibinfo  {journal} {Phys. Rev. A}\ }\textbf {\bibinfo {volume}
  {97}},\ \bibinfo {eid} {062505} (\bibinfo {year} {2018})}\BibitemShut
  {NoStop}%
\bibitem [{\citenamefont {Li}\ \emph {et~al.}(2021)\citenamefont {Li},
  \citenamefont {Qiao}, \citenamefont {Tang},\ and\ \citenamefont
  {Shi}}]{Li2021pra}%
  \BibitemOpen
  \bibfield  {author} {\bibinfo {author} {\bibfnamefont {F.-C.}\ \bibnamefont
  {Li}}, \bibinfo {author} {\bibfnamefont {H.-X.}\ \bibnamefont {Qiao}},
  \bibinfo {author} {\bibfnamefont {Y.-B.}\ \bibnamefont {Tang}}, \ and\
  \bibinfo {author} {\bibfnamefont {T.-Y.}\ \bibnamefont {Shi}},\ }\href
  {\doibase 10.1103/PhysRevA.104.062808} {\bibfield  {journal} {\bibinfo
  {journal} {Phys. Rev. A}\ }\textbf {\bibinfo {volume} {104}},\ \bibinfo
  {pages} {062808} (\bibinfo {year} {2021})}\BibitemShut {NoStop}%
\bibitem [{\citenamefont {Porsev}\ \emph {et~al.}(2021)\citenamefont {Porsev},
  \citenamefont {Safronova},\ and\ \citenamefont {Kozlov}}]{Porsev2021prl}%
  \BibitemOpen
  \bibfield  {author} {\bibinfo {author} {\bibfnamefont {S.~G.}\ \bibnamefont
  {Porsev}}, \bibinfo {author} {\bibfnamefont {M.~S.}\ \bibnamefont
  {Safronova}}, \ and\ \bibinfo {author} {\bibfnamefont {M.~G.}\ \bibnamefont
  {Kozlov}},\ }\href {\doibase 10.1103/PhysRevLett.127.253001} {\bibfield
  {journal} {\bibinfo  {journal} {Phys. Rev. Lett.}\ }\textbf {\bibinfo
  {volume} {127}},\ \bibinfo {pages} {253001} (\bibinfo {year}
  {2021})}\BibitemShut {NoStop}%
\bibitem [{\citenamefont {{Papoulia}}\ \emph {et~al.}(2021)\citenamefont
  {{Papoulia}}, \citenamefont {{Schiffmann}}, \citenamefont {{Biero{\'n}}},
  \citenamefont {{Gaigalas}}, \citenamefont {{Godefroid}}, \citenamefont
  {{Harman}}, \citenamefont {{J{\"o}nsson}}, \citenamefont {{Oreshkina}},
  \citenamefont {{Pyykk{\"o}}},\ and\ \citenamefont
  {{Tupitsyn}}}]{Papoulia2021pra}%
  \BibitemOpen
  \bibfield  {author} {\bibinfo {author} {\bibfnamefont {A.}~\bibnamefont
  {{Papoulia}}}, \bibinfo {author} {\bibfnamefont {S.}~\bibnamefont
  {{Schiffmann}}}, \bibinfo {author} {\bibfnamefont {J.}~\bibnamefont
  {{Biero{\'n}}}}, \bibinfo {author} {\bibfnamefont {G.}~\bibnamefont
  {{Gaigalas}}}, \bibinfo {author} {\bibfnamefont {M.}~\bibnamefont
  {{Godefroid}}}, \bibinfo {author} {\bibfnamefont {Z.}~\bibnamefont
  {{Harman}}}, \bibinfo {author} {\bibfnamefont {P.}~\bibnamefont
  {{J{\"o}nsson}}}, \bibinfo {author} {\bibfnamefont {N.~S.}\ \bibnamefont
  {{Oreshkina}}}, \bibinfo {author} {\bibfnamefont {P.}~\bibnamefont
  {{Pyykk{\"o}}}}, \ and\ \bibinfo {author} {\bibfnamefont {I.~I.}\
  \bibnamefont {{Tupitsyn}}},\ }\href {\doibase 10.1103/PhysRevA.103.022815}
  {\bibfield  {journal} {\bibinfo  {journal} {Phys. Rev. A}\ }\textbf {\bibinfo
  {volume} {103}},\ \bibinfo {eid} {022815} (\bibinfo {year}
  {2021})}\BibitemShut {NoStop}%
\bibitem [{\citenamefont {Skripnikov}\ \emph {et~al.}(2021)\citenamefont
  {Skripnikov}, \citenamefont {Oleynichenko}, \citenamefont {Zaitsevskii},
  \citenamefont {Maison},\ and\ \citenamefont {Barzakh}}]{Skripnikov2021prc}%
  \BibitemOpen
  \bibfield  {author} {\bibinfo {author} {\bibfnamefont {L.~V.}\ \bibnamefont
  {Skripnikov}}, \bibinfo {author} {\bibfnamefont {A.~V.}\ \bibnamefont
  {Oleynichenko}}, \bibinfo {author} {\bibfnamefont {A.~V.}\ \bibnamefont
  {Zaitsevskii}}, \bibinfo {author} {\bibfnamefont {D.~E.}\ \bibnamefont
  {Maison}}, \ and\ \bibinfo {author} {\bibfnamefont {A.~E.}\ \bibnamefont
  {Barzakh}},\ }\href {\doibase 10.1103/PhysRevC.104.034316} {\bibfield
  {journal} {\bibinfo  {journal} {Phys. Rev. C}\ }\textbf {\bibinfo {volume}
  {104}},\ \bibinfo {pages} {034316} (\bibinfo {year} {2021})}\BibitemShut
  {NoStop}%
\bibitem [{\citenamefont {Skripnikov}\ and\ \citenamefont
  {Barzakh}(2024)}]{Skripnikov2024prc}%
  \BibitemOpen
  \bibfield  {author} {\bibinfo {author} {\bibfnamefont {L.~V.}\ \bibnamefont
  {Skripnikov}}\ and\ \bibinfo {author} {\bibfnamefont {A.~E.}\ \bibnamefont
  {Barzakh}},\ }\href {\doibase 10.1103/PhysRevC.109.024315} {\bibfield
  {journal} {\bibinfo  {journal} {Phys. Rev. C}\ }\textbf {\bibinfo {volume}
  {109}},\ \bibinfo {pages} {024315} (\bibinfo {year} {2024})}\BibitemShut
  {NoStop}%
\bibitem [{\citenamefont {Lu}\ and\ \citenamefont {Chang}(2024)}]{Lu2024pra}%
  \BibitemOpen
  \bibfield  {author} {\bibinfo {author} {\bibfnamefont {B.}~\bibnamefont
  {Lu}}\ and\ \bibinfo {author} {\bibfnamefont {H.}~\bibnamefont {Chang}},\
  }\href {\doibase 10.1103/PhysRevA.110.042820} {\bibfield  {journal} {\bibinfo
   {journal} {Phys. Rev. A}\ }\textbf {\bibinfo {volume} {110}},\ \bibinfo
  {pages} {042820} (\bibinfo {year} {2024})}\BibitemShut {NoStop}%
\bibitem [{\citenamefont {Zhang}\ \emph {et~al.}(2025)\citenamefont {Zhang},
  \citenamefont {Dang}, \citenamefont {Wang},\ and\ \citenamefont
  {Tang}}]{Zhang2025pra}%
  \BibitemOpen
  \bibfield  {author} {\bibinfo {author} {\bibfnamefont {Y.-S.}\ \bibnamefont
  {Zhang}}, \bibinfo {author} {\bibfnamefont {W.}~\bibnamefont {Dang}},
  \bibinfo {author} {\bibfnamefont {K.}~\bibnamefont {Wang}}, \ and\ \bibinfo
  {author} {\bibfnamefont {Y.-B.}\ \bibnamefont {Tang}},\ }\href {\doibase
  10.1103/PhysRevA.111.042819} {\bibfield  {journal} {\bibinfo  {journal}
  {Phys. Rev. A}\ }\textbf {\bibinfo {volume} {111}},\ \bibinfo {pages}
  {042819} (\bibinfo {year} {2025})}\BibitemShut {NoStop}%
\bibitem [{\citenamefont {Tang}(2025)}]{Tang2025pra}%
  \BibitemOpen
  \bibfield  {author} {\bibinfo {author} {\bibfnamefont {Y.-B.}\ \bibnamefont
  {Tang}},\ }\href {\doibase 10.1103/l6y6-cf78} {\bibfield  {journal} {\bibinfo
   {journal} {Phys. Rev. A}\ }\textbf {\bibinfo {volume} {112}},\ \bibinfo
  {pages} {042814} (\bibinfo {year} {2025})}\BibitemShut {NoStop}%
\bibitem [{\citenamefont {{Sundholm}}\ and\ \citenamefont
  {{Olsen}}(1991)}]{Sundholm1991npa}%
  \BibitemOpen
  \bibfield  {author} {\bibinfo {author} {\bibfnamefont {D.}~\bibnamefont
  {{Sundholm}}}\ and\ \bibinfo {author} {\bibfnamefont {J.}~\bibnamefont
  {{Olsen}}},\ }\href {\doibase 10.1016/0375-9474(91)90505-Z} {\bibfield
  {journal} {\bibinfo  {journal} {Nucl. Phys. A}\ }\textbf {\bibinfo {volume}
  {534}},\ \bibinfo {pages} {360} (\bibinfo {year} {1991})}\BibitemShut
  {NoStop}%
\bibitem [{\citenamefont {{Weber}}\ \emph {et~al.}(1982)\citenamefont
  {{Weber}}, \citenamefont {{Jeckelmann}}, \citenamefont {{Kern}},
  \citenamefont {{Kiebele}}, \citenamefont {{Aas}}, \citenamefont {{Beer}},
  \citenamefont {{Beltrami}}, \citenamefont {{Bos}}, \citenamefont {{de
  Chambrier}}, \citenamefont {{Goudsmit}}, \citenamefont {{Leisi}},
  \citenamefont {{Ruckstuhl}}, \citenamefont {{Strassner}},\ and\ \citenamefont
  {{Vacchi}}}]{Weber1982npa}%
  \BibitemOpen
  \bibfield  {author} {\bibinfo {author} {\bibfnamefont {R.}~\bibnamefont
  {{Weber}}}, \bibinfo {author} {\bibfnamefont {B.}~\bibnamefont
  {{Jeckelmann}}}, \bibinfo {author} {\bibfnamefont {J.}~\bibnamefont
  {{Kern}}}, \bibinfo {author} {\bibfnamefont {U.}~\bibnamefont {{Kiebele}}},
  \bibinfo {author} {\bibfnamefont {B.}~\bibnamefont {{Aas}}}, \bibinfo
  {author} {\bibfnamefont {W.}~\bibnamefont {{Beer}}}, \bibinfo {author}
  {\bibfnamefont {I.}~\bibnamefont {{Beltrami}}}, \bibinfo {author}
  {\bibfnamefont {K.}~\bibnamefont {{Bos}}}, \bibinfo {author} {\bibfnamefont
  {G.}~\bibnamefont {{de Chambrier}}}, \bibinfo {author} {\bibfnamefont
  {P.~F.~A.}\ \bibnamefont {{Goudsmit}}}, \bibinfo {author} {\bibfnamefont
  {H.~J.}\ \bibnamefont {{Leisi}}}, \bibinfo {author} {\bibfnamefont
  {W.}~\bibnamefont {{Ruckstuhl}}}, \bibinfo {author} {\bibfnamefont
  {G.}~\bibnamefont {{Strassner}}}, \ and\ \bibinfo {author} {\bibfnamefont
  {A.}~\bibnamefont {{Vacchi}}},\ }\href {\doibase
  10.1016/0375-9474(82)90046-X} {\bibfield  {journal} {\bibinfo  {journal}
  {Nucl. Phys. A}\ }\textbf {\bibinfo {volume} {377}},\ \bibinfo {pages} {361}
  (\bibinfo {year} {1982})}\BibitemShut {NoStop}%
\bibitem [{\citenamefont {Barwood}\ \emph {et~al.}(2003)\citenamefont
  {Barwood}, \citenamefont {Gao}, \citenamefont {Gill}, \citenamefont {Huang},\
  and\ \citenamefont {Klein}}]{Barwood2003pra}%
  \BibitemOpen
  \bibfield  {author} {\bibinfo {author} {\bibfnamefont {G.~P.}\ \bibnamefont
  {Barwood}}, \bibinfo {author} {\bibfnamefont {K.}~\bibnamefont {Gao}},
  \bibinfo {author} {\bibfnamefont {P.}~\bibnamefont {Gill}}, \bibinfo {author}
  {\bibfnamefont {G.}~\bibnamefont {Huang}}, \ and\ \bibinfo {author}
  {\bibfnamefont {H.~A.}\ \bibnamefont {Klein}},\ }\href {\doibase
  10.1103/PhysRevA.67.013402} {\bibfield  {journal} {\bibinfo  {journal} {Phys.
  Rev. A}\ }\textbf {\bibinfo {volume} {67}},\ \bibinfo {pages} {013402}
  (\bibinfo {year} {2003})}\BibitemShut {NoStop}%
\bibitem [{\citenamefont {Sahoo}(2006{\natexlab{a}})}]{Sahoo2006pra}%
  \BibitemOpen
  \bibfield  {author} {\bibinfo {author} {\bibfnamefont {B.~K.}\ \bibnamefont
  {Sahoo}},\ }\href {\doibase 10.1103/PhysRevA.73.062501} {\bibfield  {journal}
  {\bibinfo  {journal} {Phys. Rev. A}\ }\textbf {\bibinfo {volume} {73}},\
  \bibinfo {pages} {062501} (\bibinfo {year} {2006}{\natexlab{a}})}\BibitemShut
  {NoStop}%
\bibitem [{\citenamefont {Lu}\ \emph {et~al.}(2019)\citenamefont {Lu},
  \citenamefont {Zhang}, \citenamefont {Chang}, \citenamefont {Li},
  \citenamefont {Wu},\ and\ \citenamefont {Wang}}]{Lu2019pra}%
  \BibitemOpen
  \bibfield  {author} {\bibinfo {author} {\bibfnamefont {B.}~\bibnamefont
  {Lu}}, \bibinfo {author} {\bibfnamefont {T.}~\bibnamefont {Zhang}}, \bibinfo
  {author} {\bibfnamefont {H.}~\bibnamefont {Chang}}, \bibinfo {author}
  {\bibfnamefont {J.}~\bibnamefont {Li}}, \bibinfo {author} {\bibfnamefont
  {Y.}~\bibnamefont {Wu}}, \ and\ \bibinfo {author} {\bibfnamefont
  {J.}~\bibnamefont {Wang}},\ }\href {\doibase 10.1103/PhysRevA.100.012504}
  {\bibfield  {journal} {\bibinfo  {journal} {Phys. Rev. A}\ }\textbf {\bibinfo
  {volume} {100}},\ \bibinfo {pages} {012504} (\bibinfo {year}
  {2019})}\BibitemShut {NoStop}%
\bibitem [{\citenamefont {Silverans}\ \emph {et~al.}(1986)\citenamefont
  {Silverans}, \citenamefont {Borghs}, \citenamefont {De~Bisschop},\ and\
  \citenamefont {Van~Hove}}]{Silverans1986pra}%
  \BibitemOpen
  \bibfield  {author} {\bibinfo {author} {\bibfnamefont {R.~E.}\ \bibnamefont
  {Silverans}}, \bibinfo {author} {\bibfnamefont {G.}~\bibnamefont {Borghs}},
  \bibinfo {author} {\bibfnamefont {P.}~\bibnamefont {De~Bisschop}}, \ and\
  \bibinfo {author} {\bibfnamefont {M.}~\bibnamefont {Van~Hove}},\ }\href
  {\doibase 10.1103/PhysRevA.33.2117} {\bibfield  {journal} {\bibinfo
  {journal} {Phys. Rev. A}\ }\textbf {\bibinfo {volume} {33}},\ \bibinfo
  {pages} {2117} (\bibinfo {year} {1986})}\BibitemShut {NoStop}%
\bibitem [{\citenamefont {{Wendt}}\ \emph {et~al.}(1988)\citenamefont
  {{Wendt}}, \citenamefont {{Ahmad}}, \citenamefont {{Ekstr{\"o}m}},
  \citenamefont {{Klempt}}, \citenamefont {{Neugart}},\ and\ \citenamefont
  {{Otten}}}]{Wendt1988ZPA}%
  \BibitemOpen
  \bibfield  {author} {\bibinfo {author} {\bibfnamefont {K.}~\bibnamefont
  {{Wendt}}}, \bibinfo {author} {\bibfnamefont {S.~A.}\ \bibnamefont
  {{Ahmad}}}, \bibinfo {author} {\bibfnamefont {C.}~\bibnamefont
  {{Ekstr{\"o}m}}}, \bibinfo {author} {\bibfnamefont {W.}~\bibnamefont
  {{Klempt}}}, \bibinfo {author} {\bibfnamefont {R.}~\bibnamefont {{Neugart}}},
  \ and\ \bibinfo {author} {\bibfnamefont {E.~W.}\ \bibnamefont {{Otten}}},\
  }\href {\doibase 10.1007/BF01294345} {\bibfield  {journal} {\bibinfo
  {journal} {Z. Phys. A}\ }\textbf {\bibinfo {volume} {329}},\ \bibinfo {pages}
  {407} (\bibinfo {year} {1988})}\BibitemShut {NoStop}%
\bibitem [{\citenamefont {Lurio}(1962)}]{LurioPR1961}%
  \BibitemOpen
  \bibfield  {author} {\bibinfo {author} {\bibfnamefont {A.}~\bibnamefont
  {Lurio}},\ }\href {\doibase 10.1103/PhysRev.126.1768} {\bibfield  {journal}
  {\bibinfo  {journal} {Phys. Rev.}\ }\textbf {\bibinfo {volume} {126}},\
  \bibinfo {pages} {1768} (\bibinfo {year} {1962})}\BibitemShut {NoStop}%
\bibitem [{\citenamefont {{Grundevik}}\ \emph {et~al.}(1983)\citenamefont
  {{Grundevik}}, \citenamefont {{Gustavsson}}, \citenamefont {{Lindgren}},
  \citenamefont {{Olsson}}, \citenamefont {{Olsson}},\ and\ \citenamefont
  {{Ros{\'e}n}}}]{GrundevikZPA1983}%
  \BibitemOpen
  \bibfield  {author} {\bibinfo {author} {\bibfnamefont {P.}~\bibnamefont
  {{Grundevik}}}, \bibinfo {author} {\bibfnamefont {M.}~\bibnamefont
  {{Gustavsson}}}, \bibinfo {author} {\bibfnamefont {I.}~\bibnamefont
  {{Lindgren}}}, \bibinfo {author} {\bibfnamefont {G.}~\bibnamefont
  {{Olsson}}}, \bibinfo {author} {\bibfnamefont {T.}~\bibnamefont {{Olsson}}},
  \ and\ \bibinfo {author} {\bibfnamefont {A.}~\bibnamefont {{Ros{\'e}n}}},\
  }\href {\doibase 10.1007/BF01415098} {\bibfield  {journal} {\bibinfo
  {journal} {Z. Phys. A}\ }\textbf {\bibinfo {volume} {311}},\ \bibinfo {pages}
  {143} (\bibinfo {year} {1983})}\BibitemShut {NoStop}%
\bibitem [{\citenamefont {{Zu Putlitz}}(1963)}]{PutlitzZP1963}%
  \BibitemOpen
  \bibfield  {author} {\bibinfo {author} {\bibfnamefont {G.}~\bibnamefont {{Zu
  Putlitz}}},\ }\href {\doibase 10.1007/BF01375346} {\bibfield  {journal}
  {\bibinfo  {journal} {Z. Phys.}\ }\textbf {\bibinfo {volume} {175}},\
  \bibinfo {pages} {543} (\bibinfo {year} {1963})}\BibitemShut {NoStop}%
\bibitem [{\citenamefont {Heider}\ and\ \citenamefont
  {Brink}(1977)}]{HeiderPRA1977}%
  \BibitemOpen
  \bibfield  {author} {\bibinfo {author} {\bibfnamefont {S.~M.}\ \bibnamefont
  {Heider}}\ and\ \bibinfo {author} {\bibfnamefont {G.~O.}\ \bibnamefont
  {Brink}},\ }\href {\doibase 10.1103/PhysRevA.16.1371} {\bibfield  {journal}
  {\bibinfo  {journal} {Phys. Rev. A}\ }\textbf {\bibinfo {volume} {16}},\
  \bibinfo {pages} {1371} (\bibinfo {year} {1977})}\BibitemShut {NoStop}%
\bibitem [{\citenamefont {{Gustavsson}}\ \emph {et~al.}(1979)\citenamefont
  {{Gustavsson}}, \citenamefont {{Olsson}},\ and\ \citenamefont
  {{Ros{\'e}n}}}]{GustavssonZPA1979}%
  \BibitemOpen
  \bibfield  {author} {\bibinfo {author} {\bibfnamefont {M.}~\bibnamefont
  {{Gustavsson}}}, \bibinfo {author} {\bibfnamefont {G.}~\bibnamefont
  {{Olsson}}}, \ and\ \bibinfo {author} {\bibfnamefont {A.}~\bibnamefont
  {{Ros{\'e}n}}},\ }\href {\doibase 10.1007/BF01408538} {\bibfield  {journal}
  {\bibinfo  {journal} {Z. Phys. A}\ }\textbf {\bibinfo {volume} {290}},\
  \bibinfo {pages} {231} (\bibinfo {year} {1979})}\BibitemShut {NoStop}%
\bibitem [{\citenamefont {{Schmelling}}(1974)}]{SchmellingPRA1974}%
  \BibitemOpen
  \bibfield  {author} {\bibinfo {author} {\bibfnamefont {S.~G.}\ \bibnamefont
  {{Schmelling}}},\ }\href {\doibase 10.1103/PhysRevA.9.1097} {\bibfield
  {journal} {\bibinfo  {journal} {Phys. Rev. A}\ }\textbf {\bibinfo {volume}
  {9}},\ \bibinfo {pages} {1097} (\bibinfo {year} {1974})}\BibitemShut
  {NoStop}%
\bibitem [{\citenamefont {{Wijngaarden}}\ and\ \citenamefont
  {{Li}}(1995)}]{vanCJP1995}%
  \BibitemOpen
  \bibfield  {author} {\bibinfo {author} {\bibfnamefont {W.~A.~v.}\
  \bibnamefont {{Wijngaarden}}}\ and\ \bibinfo {author} {\bibfnamefont
  {J.}~\bibnamefont {{Li}}},\ }\href {\doibase 10.1139/p95-069} {\bibfield
  {journal} {\bibinfo  {journal} {Can. J. Phys.}\ }\textbf {\bibinfo {volume}
  {73}},\ \bibinfo {pages} {484} (\bibinfo {year} {1995})}\BibitemShut
  {NoStop}%
\bibitem [{\citenamefont {Safronova}\ \emph {et~al.}(2009)\citenamefont
  {Safronova}, \citenamefont {Kozlov}, \citenamefont {Johnson},\ and\
  \citenamefont {Jiang}}]{Safronova2009PRA}%
  \BibitemOpen
  \bibfield  {author} {\bibinfo {author} {\bibfnamefont {M.~S.}\ \bibnamefont
  {Safronova}}, \bibinfo {author} {\bibfnamefont {M.~G.}\ \bibnamefont
  {Kozlov}}, \bibinfo {author} {\bibfnamefont {W.~R.}\ \bibnamefont {Johnson}},
  \ and\ \bibinfo {author} {\bibfnamefont {D.}~\bibnamefont {Jiang}},\ }\href
  {\doibase 10.1103/PhysRevA.80.012516} {\bibfield  {journal} {\bibinfo
  {journal} {Phys. Rev. A}\ }\textbf {\bibinfo {volume} {80}},\ \bibinfo
  {pages} {012516} (\bibinfo {year} {2009})}\BibitemShut {NoStop}%
\bibitem [{\citenamefont {Dzuba}(2014)}]{Dzuba2014PRA}%
  \BibitemOpen
  \bibfield  {author} {\bibinfo {author} {\bibfnamefont {V.~A.}\ \bibnamefont
  {Dzuba}},\ }\href {\doibase 10.1103/PhysRevA.90.012517} {\bibfield  {journal}
  {\bibinfo  {journal} {Phys. Rev. A}\ }\textbf {\bibinfo {volume} {90}},\
  \bibinfo {pages} {012517} (\bibinfo {year} {2014})}\BibitemShut {NoStop}%
\bibitem [{\citenamefont {Dzuba}\ \emph {et~al.}(1996)\citenamefont {Dzuba},
  \citenamefont {Flambaum},\ and\ \citenamefont {Kozlov}}]{Dzuba1996}%
  \BibitemOpen
  \bibfield  {author} {\bibinfo {author} {\bibfnamefont {V.~A.}\ \bibnamefont
  {Dzuba}}, \bibinfo {author} {\bibfnamefont {V.~V.}\ \bibnamefont {Flambaum}},
  \ and\ \bibinfo {author} {\bibfnamefont {M.~G.}\ \bibnamefont {Kozlov}},\
  }\href {\doibase 10.1103/PhysRevA.54.3948} {\bibfield  {journal} {\bibinfo
  {journal} {Phys. Rev. A}\ }\textbf {\bibinfo {volume} {54}},\ \bibinfo
  {pages} {3948} (\bibinfo {year} {1996})}\BibitemShut {NoStop}%
\bibitem [{\citenamefont {Tang}\ \emph {et~al.}(2017)\citenamefont {Tang},
  \citenamefont {Lou},\ and\ \citenamefont {Shi}}]{Tang2017PRA}%
  \BibitemOpen
  \bibfield  {author} {\bibinfo {author} {\bibfnamefont {Y.-B.}\ \bibnamefont
  {Tang}}, \bibinfo {author} {\bibfnamefont {B.-Q.}\ \bibnamefont {Lou}}, \
  and\ \bibinfo {author} {\bibfnamefont {T.-Y.}\ \bibnamefont {Shi}},\ }\href
  {\doibase 10.1103/PhysRevA.96.022513} {\bibfield  {journal} {\bibinfo
  {journal} {Phys. Rev. A}\ }\textbf {\bibinfo {volume} {96}},\ \bibinfo
  {pages} {022513} (\bibinfo {year} {2017})}\BibitemShut {NoStop}%
\bibitem [{\citenamefont {Blundell}\ \emph {et~al.}(1989)\citenamefont
  {Blundell}, \citenamefont {Johnson}, \citenamefont {Liu},\ and\ \citenamefont
  {Sapirstein}}]{Blundell1989PRA}%
  \BibitemOpen
  \bibfield  {author} {\bibinfo {author} {\bibfnamefont {S.~A.}\ \bibnamefont
  {Blundell}}, \bibinfo {author} {\bibfnamefont {W.~R.}\ \bibnamefont
  {Johnson}}, \bibinfo {author} {\bibfnamefont {Z.~W.}\ \bibnamefont {Liu}}, \
  and\ \bibinfo {author} {\bibfnamefont {J.}~\bibnamefont {Sapirstein}},\
  }\href {\doibase 10.1103/PhysRevA.40.2233} {\bibfield  {journal} {\bibinfo
  {journal} {Phys. Rev. A}\ }\textbf {\bibinfo {volume} {40}},\ \bibinfo
  {pages} {2233} (\bibinfo {year} {1989})}\BibitemShut {NoStop}%
\bibitem [{\citenamefont {{Dzuba}}\ \emph {et~al.}(1998)\citenamefont
  {{Dzuba}}, \citenamefont {{Flambaum}}, \citenamefont {{Kozlov}},\ and\
  \citenamefont {{Porsev}}}]{Dzuba1998JETP}%
  \BibitemOpen
  \bibfield  {author} {\bibinfo {author} {\bibfnamefont {V.~A.}\ \bibnamefont
  {{Dzuba}}}, \bibinfo {author} {\bibfnamefont {V.~V.}\ \bibnamefont
  {{Flambaum}}}, \bibinfo {author} {\bibfnamefont {M.~G.}\ \bibnamefont
  {{Kozlov}}}, \ and\ \bibinfo {author} {\bibfnamefont {S.~G.}\ \bibnamefont
  {{Porsev}}},\ }\href {\doibase 10.1134/1.558736} {\bibfield  {journal}
  {\bibinfo  {journal} {J. Exp. Theor. Phys.}\ }\textbf {\bibinfo {volume}
  {87}},\ \bibinfo {pages} {885} (\bibinfo {year} {1998})}\BibitemShut
  {NoStop}%
\bibitem [{\citenamefont {{Safronova}}\ \emph {et~al.}(1999)\citenamefont
  {{Safronova}}, \citenamefont {{Derevianko}}, \citenamefont {{Safronova}},\
  and\ \citenamefont {{Johnson}}}]{Safronova1999JPB}%
  \BibitemOpen
  \bibfield  {author} {\bibinfo {author} {\bibfnamefont {U.~I.}\ \bibnamefont
  {{Safronova}}}, \bibinfo {author} {\bibfnamefont {A.}~\bibnamefont
  {{Derevianko}}}, \bibinfo {author} {\bibfnamefont {M.~S.}\ \bibnamefont
  {{Safronova}}}, \ and\ \bibinfo {author} {\bibfnamefont {W.~R.}\ \bibnamefont
  {{Johnson}}},\ }\href {\doibase 10.1088/0953-4075/32/14/319} {\bibfield
  {journal} {\bibinfo  {journal} {J. Phys. B: At. Mol. Phys.}\ }\textbf
  {\bibinfo {volume} {32}},\ \bibinfo {pages} {3527} (\bibinfo {year}
  {1999})}\BibitemShut {NoStop}%
\bibitem [{\citenamefont {Savukov}(2004)}]{Savukov2004PRA}%
  \BibitemOpen
  \bibfield  {author} {\bibinfo {author} {\bibfnamefont {I.~M.}\ \bibnamefont
  {Savukov}},\ }\href {\doibase 10.1103/PhysRevA.70.042502} {\bibfield
  {journal} {\bibinfo  {journal} {Phys. Rev. A}\ }\textbf {\bibinfo {volume}
  {70}},\ \bibinfo {pages} {042502} (\bibinfo {year} {2004})}\BibitemShut
  {NoStop}%
\bibitem [{\citenamefont {Chaudhuri}\ \emph {et~al.}(1999)\citenamefont
  {Chaudhuri}, \citenamefont {Panda},\ and\ \citenamefont
  {Das}}]{Chaudhuri1999pra}%
  \BibitemOpen
  \bibfield  {author} {\bibinfo {author} {\bibfnamefont {R.~K.}\ \bibnamefont
  {Chaudhuri}}, \bibinfo {author} {\bibfnamefont {P.~K.}\ \bibnamefont
  {Panda}}, \ and\ \bibinfo {author} {\bibfnamefont {B.~P.}\ \bibnamefont
  {Das}},\ }\href {\doibase 10.1103/PhysRevA.59.1187} {\bibfield  {journal}
  {\bibinfo  {journal} {Phys. Rev. A}\ }\textbf {\bibinfo {volume} {59}},\
  \bibinfo {pages} {1187} (\bibinfo {year} {1999})}\BibitemShut {NoStop}%
\bibitem [{\citenamefont {Zhang}\ \emph {et~al.}(2023)\citenamefont {Zhang},
  \citenamefont {Jiang}, \citenamefont {Dong},\ and\ \citenamefont
  {Tang}}]{Zhang2023PRA}%
  \BibitemOpen
  \bibfield  {author} {\bibinfo {author} {\bibfnamefont {R.-K.}\ \bibnamefont
  {Zhang}}, \bibinfo {author} {\bibfnamefont {J.}~\bibnamefont {Jiang}},
  \bibinfo {author} {\bibfnamefont {C.-Z.}\ \bibnamefont {Dong}}, \ and\
  \bibinfo {author} {\bibfnamefont {Y.-B.}\ \bibnamefont {Tang}},\ }\href
  {\doibase 10.1103/PhysRevA.108.012803} {\bibfield  {journal} {\bibinfo
  {journal} {Phys. Rev. A}\ }\textbf {\bibinfo {volume} {108}},\ \bibinfo
  {pages} {012803} (\bibinfo {year} {2023})}\BibitemShut {NoStop}%
\bibitem [{\citenamefont {Zhang}\ \emph {et~al.}(2024)\citenamefont {Zhang},
  \citenamefont {Jiang}, \citenamefont {Dong},\ and\ \citenamefont
  {Tang}}]{Zhang2024PRA}%
  \BibitemOpen
  \bibfield  {author} {\bibinfo {author} {\bibfnamefont {R.-K.}\ \bibnamefont
  {Zhang}}, \bibinfo {author} {\bibfnamefont {J.}~\bibnamefont {Jiang}},
  \bibinfo {author} {\bibfnamefont {C.-Z.}\ \bibnamefont {Dong}}, \ and\
  \bibinfo {author} {\bibfnamefont {Y.-B.}\ \bibnamefont {Tang}},\ }\href
  {\doibase 10.1103/PhysRevA.109.032821} {\bibfield  {journal} {\bibinfo
  {journal} {Phys. Rev. A}\ }\textbf {\bibinfo {volume} {109}},\ \bibinfo
  {pages} {032821} (\bibinfo {year} {2024})}\BibitemShut {NoStop}%
\bibitem [{\citenamefont {Kramida}\ \emph {et~al.}(2024)\citenamefont
  {Kramida}, \citenamefont {{Yu.~Ralchenko}}, \citenamefont {Reader},\ and\
  \citenamefont {{and NIST ASD Team}}}]{NIST_ASD}%
  \BibitemOpen
  \bibfield  {author} {\bibinfo {author} {\bibfnamefont {A.}~\bibnamefont
  {Kramida}}, \bibinfo {author} {\bibnamefont {{Yu.~Ralchenko}}}, \bibinfo
  {author} {\bibfnamefont {J.}~\bibnamefont {Reader}}, \ and\ \bibinfo {author}
  {\bibnamefont {{and NIST ASD Team}}},\ }\href@noop {} {}\bibinfo
  {howpublished} {{NIST Atomic Spectra Database (ver. 5.12), [Online].
  Available: {doi:https://doi.org/10.18434/T4W30F}. National Institute of
  Standards and Technology, Gaithersburg, MD.}} (\bibinfo {year}
  {2024})\BibitemShut {NoStop}%
\bibitem [{\citenamefont {{Stone}}(2005)}]{Stone2005ADNDT}%
  \BibitemOpen
  \bibfield  {author} {\bibinfo {author} {\bibfnamefont {N.~J.}\ \bibnamefont
  {{Stone}}},\ }\href {\doibase 10.1016/j.adt.2005.04.001} {\bibfield
  {journal} {\bibinfo  {journal} {At. Data Nucl. Data Tables}\ }\textbf
  {\bibinfo {volume} {90}},\ \bibinfo {pages} {75} (\bibinfo {year}
  {2005})}\BibitemShut {NoStop}%
\bibitem [{\citenamefont {He}\ \emph {et~al.}(2009)\citenamefont {He},
  \citenamefont {Therkildsen}, \citenamefont {Jensen}, \citenamefont {Brusch},
  \citenamefont {Thomsen},\ and\ \citenamefont {Porsev}}]{HePRA2009}%
  \BibitemOpen
  \bibfield  {author} {\bibinfo {author} {\bibfnamefont {M.}~\bibnamefont
  {He}}, \bibinfo {author} {\bibfnamefont {K.~T.}\ \bibnamefont {Therkildsen}},
  \bibinfo {author} {\bibfnamefont {B.~B.}\ \bibnamefont {Jensen}}, \bibinfo
  {author} {\bibfnamefont {A.}~\bibnamefont {Brusch}}, \bibinfo {author}
  {\bibfnamefont {J.~W.}\ \bibnamefont {Thomsen}}, \ and\ \bibinfo {author}
  {\bibfnamefont {S.~G.}\ \bibnamefont {Porsev}},\ }\href {\doibase
  10.1103/PhysRevA.80.024501} {\bibfield  {journal} {\bibinfo  {journal} {Phys.
  Rev. A}\ }\textbf {\bibinfo {volume} {80}},\ \bibinfo {pages} {024501}
  (\bibinfo {year} {2009})}\BibitemShut {NoStop}%
\bibitem [{\citenamefont {{Veseth}}(1987)}]{Veseth1987JPhB}%
  \BibitemOpen
  \bibfield  {author} {\bibinfo {author} {\bibfnamefont {L.}~\bibnamefont
  {{Veseth}}},\ }\href {\doibase 10.1088/0022-3700/20/2/009} {\bibfield
  {journal} {\bibinfo  {journal} {J. Phys. B: At. Mol. Phys.}\ }\textbf
  {\bibinfo {volume} {20}},\ \bibinfo {pages} {235} (\bibinfo {year}
  {1987})}\BibitemShut {NoStop}%
\bibitem [{\citenamefont {Porsev}\ and\ \citenamefont
  {Derevianko}(2004)}]{Porsev2004PRA}%
  \BibitemOpen
  \bibfield  {author} {\bibinfo {author} {\bibfnamefont {S.~G.}\ \bibnamefont
  {Porsev}}\ and\ \bibinfo {author} {\bibfnamefont {A.}~\bibnamefont
  {Derevianko}},\ }\href {\doibase 10.1103/PhysRevA.69.042506} {\bibfield
  {journal} {\bibinfo  {journal} {Phys. Rev. A}\ }\textbf {\bibinfo {volume}
  {69}},\ \bibinfo {pages} {042506} (\bibinfo {year} {2004})}\BibitemShut
  {NoStop}%
\bibitem [{\citenamefont {Beloy}\ \emph {et~al.}(2008)\citenamefont {Beloy},
  \citenamefont {Derevianko},\ and\ \citenamefont {Johnson}}]{Beloy2008PRA}%
  \BibitemOpen
  \bibfield  {author} {\bibinfo {author} {\bibfnamefont {K.}~\bibnamefont
  {Beloy}}, \bibinfo {author} {\bibfnamefont {A.}~\bibnamefont {Derevianko}}, \
  and\ \bibinfo {author} {\bibfnamefont {W.~R.}\ \bibnamefont {Johnson}},\
  }\href {\doibase 10.1103/PhysRevA.77.012512} {\bibfield  {journal} {\bibinfo
  {journal} {Phys. Rev. A}\ }\textbf {\bibinfo {volume} {77}},\ \bibinfo
  {pages} {012512} (\bibinfo {year} {2008})}\BibitemShut {NoStop}%
\bibitem [{\citenamefont {{Kozlov}}\ and\ \citenamefont
  {{Porsev}}(1999)}]{Kozlov1999ejpd}%
  \BibitemOpen
  \bibfield  {author} {\bibinfo {author} {\bibfnamefont {M.~G.}\ \bibnamefont
  {{Kozlov}}}\ and\ \bibinfo {author} {\bibfnamefont {S.~G.}\ \bibnamefont
  {{Porsev}}},\ }\href {\doibase 10.1007/s100530050229} {\bibfield  {journal}
  {\bibinfo  {journal} {Eur. Phys. J. D}\ }\textbf {\bibinfo {volume} {5}},\
  \bibinfo {pages} {59} (\bibinfo {year} {1999})}\BibitemShut {NoStop}%
\bibitem [{\citenamefont {{Lutz}}\ and\ \citenamefont
  {{Oehler}}(1978)}]{Lutz1978ZPA}%
  \BibitemOpen
  \bibfield  {author} {\bibinfo {author} {\bibfnamefont {O.}~\bibnamefont
  {{Lutz}}}\ and\ \bibinfo {author} {\bibfnamefont {H.}~\bibnamefont
  {{Oehler}}},\ }\href {\doibase 10.1007/BF01408194} {\bibfield  {journal}
  {\bibinfo  {journal} {Z. Phys. A}\ }\textbf {\bibinfo {volume} {288}},\
  \bibinfo {pages} {11} (\bibinfo {year} {1978})}\BibitemShut {NoStop}%
\bibitem [{\citenamefont {{Schwalm}}\ \emph {et~al.}(1977)\citenamefont
  {{Schwalm}}, \citenamefont {{Warburton}},\ and\ \citenamefont
  {{Olness}}}]{Schwalm1977npa}%
  \BibitemOpen
  \bibfield  {author} {\bibinfo {author} {\bibfnamefont {D.}~\bibnamefont
  {{Schwalm}}}, \bibinfo {author} {\bibfnamefont {E.~K.}\ \bibnamefont
  {{Warburton}}}, \ and\ \bibinfo {author} {\bibfnamefont {J.~W.}\ \bibnamefont
  {{Olness}}},\ }\href {\doibase 10.1016/0375-9474(77)90108-7} {\bibfield
  {journal} {\bibinfo  {journal} {Nucl. Phys. A}\ }\textbf {\bibinfo {volume}
  {293}},\ \bibinfo {pages} {425} (\bibinfo {year} {1977})}\BibitemShut
  {NoStop}%
\bibitem [{\citenamefont
  {{M{\r{a}}rtensson-Pendrill}}(2002)}]{Pendrill2002JPhB}%
  \BibitemOpen
  \bibfield  {author} {\bibinfo {author} {\bibfnamefont {A.-M.}\ \bibnamefont
  {{M{\r{a}}rtensson-Pendrill}}},\ }\href {\doibase 10.1088/0953-4075/35/4/315}
  {\bibfield  {journal} {\bibinfo  {journal} {J. Phys. B: At. Mol. Phys.}\
  }\textbf {\bibinfo {volume} {35}},\ \bibinfo {pages} {917} (\bibinfo {year}
  {2002})}\BibitemShut {NoStop}%
\bibitem [{\citenamefont {Yu}\ \emph {et~al.}(2004)\citenamefont {Yu},
  \citenamefont {Wu}, \citenamefont {Gou},\ and\ \citenamefont
  {Shi}}]{Yu2004PRA}%
  \BibitemOpen
  \bibfield  {author} {\bibinfo {author} {\bibfnamefont {K.-Z.}\ \bibnamefont
  {Yu}}, \bibinfo {author} {\bibfnamefont {L.-J.}\ \bibnamefont {Wu}}, \bibinfo
  {author} {\bibfnamefont {B.-C.}\ \bibnamefont {Gou}}, \ and\ \bibinfo
  {author} {\bibfnamefont {T.-Y.}\ \bibnamefont {Shi}},\ }\href {\doibase
  10.1103/PhysRevA.70.012506} {\bibfield  {journal} {\bibinfo  {journal} {Phys.
  Rev. A}\ }\textbf {\bibinfo {volume} {70}},\ \bibinfo {pages} {012506}
  (\bibinfo {year} {2004})}\BibitemShut {NoStop}%
\bibitem [{\citenamefont {Chakraborty}\ \emph {et~al.}(2026)\citenamefont
  {Chakraborty}, \citenamefont {Katyal},\ and\ \citenamefont
  {Sahoo}}]{Chakraborty2025arxiv}%
  \BibitemOpen
  \bibfield  {author} {\bibinfo {author} {\bibfnamefont {A.}~\bibnamefont
  {Chakraborty}}, \bibinfo {author} {\bibfnamefont {V.}~\bibnamefont {Katyal}},
  \ and\ \bibinfo {author} {\bibfnamefont {B.~K.}\ \bibnamefont {Sahoo}},\
  }\href {\doibase 10.1103/zh99-b1m3} {\bibfield  {journal} {\bibinfo
  {journal} {Phys. Rev. A}\ }\textbf {\bibinfo {volume} {113}},\ \bibinfo
  {pages} {L011101} (\bibinfo {year} {2026})}\BibitemShut {NoStop}%
\bibitem [{\citenamefont {Sahoo}(2006{\natexlab{b}})}]{Sahoo2006PRAa}%
  \BibitemOpen
  \bibfield  {author} {\bibinfo {author} {\bibfnamefont {B.~K.}\ \bibnamefont
  {Sahoo}},\ }\href {\doibase 10.1103/PhysRevA.74.020501} {\bibfield  {journal}
  {\bibinfo  {journal} {Phys. Rev. A}\ }\textbf {\bibinfo {volume} {74}},\
  \bibinfo {pages} {020501} (\bibinfo {year} {2006}{\natexlab{b}})}\BibitemShut
  {NoStop}%
\bibitem [{\citenamefont {{Jackson}}\ and\ \citenamefont
  {{Tuan}}(1964)}]{Jackson964rsa}%
  \BibitemOpen
  \bibfield  {author} {\bibinfo {author} {\bibfnamefont {D.~A.}\ \bibnamefont
  {{Jackson}}}\ and\ \bibinfo {author} {\bibfnamefont {D.~H.}\ \bibnamefont
  {{Tuan}}},\ }\href {\doibase 10.1098/rspa.1964.0148} {\bibfield  {journal}
  {\bibinfo  {journal} {Proc. R. Soc. Lond. A}\ }\textbf {\bibinfo {volume}
  {280}},\ \bibinfo {pages} {323} (\bibinfo {year} {1964})}\BibitemShut
  {NoStop}%
\bibitem [{\citenamefont {{Grundevik}}\ \emph {et~al.}(1982)\citenamefont
  {{Grundevik}}, \citenamefont {{Lundberg}}, \citenamefont {{Nilsson}},\ and\
  \citenamefont {{Olsson}}}]{Grundevik1982zpa}%
  \BibitemOpen
  \bibfield  {author} {\bibinfo {author} {\bibfnamefont {P.}~\bibnamefont
  {{Grundevik}}}, \bibinfo {author} {\bibfnamefont {H.}~\bibnamefont
  {{Lundberg}}}, \bibinfo {author} {\bibfnamefont {L.}~\bibnamefont
  {{Nilsson}}}, \ and\ \bibinfo {author} {\bibfnamefont {G.}~\bibnamefont
  {{Olsson}}},\ }\href {\doibase 10.1007/BF01415122} {\bibfield  {journal}
  {\bibinfo  {journal} {Z. Phys. A}\ }\textbf {\bibinfo {volume} {306}},\
  \bibinfo {pages} {195} (\bibinfo {year} {1982})}\BibitemShut {NoStop}%
\bibitem [{\citenamefont {{Villemoes}}\ \emph {et~al.}(1993)\citenamefont
  {{Villemoes}}, \citenamefont {{Arnesen}}, \citenamefont {{Heijkenskjold}},\
  and\ \citenamefont {{Wannstrom}}}]{Villemoes1993jpb}%
  \BibitemOpen
  \bibfield  {author} {\bibinfo {author} {\bibfnamefont {P.}~\bibnamefont
  {{Villemoes}}}, \bibinfo {author} {\bibfnamefont {A.}~\bibnamefont
  {{Arnesen}}}, \bibinfo {author} {\bibfnamefont {F.}~\bibnamefont
  {{Heijkenskjold}}}, \ and\ \bibinfo {author} {\bibfnamefont {A.}~\bibnamefont
  {{Wannstrom}}},\ }\href {\doibase 10.1088/0953-4075/26/22/030} {\bibfield
  {journal} {\bibinfo  {journal} {J. Phys. B: At. Mol. Phys.}\ }\textbf
  {\bibinfo {volume} {26}},\ \bibinfo {pages} {4289} (\bibinfo {year}
  {1993})}\BibitemShut {NoStop}%
\bibitem [{\citenamefont {Lewty}\ \emph {et~al.}(2013)\citenamefont {Lewty},
  \citenamefont {Chuah}, \citenamefont {Cazan}, \citenamefont {Barrett},\ and\
  \citenamefont {Sahoo}}]{Lewty_2013_pra}%
  \BibitemOpen
  \bibfield  {author} {\bibinfo {author} {\bibfnamefont {N.~C.}\ \bibnamefont
  {Lewty}}, \bibinfo {author} {\bibfnamefont {B.~L.}\ \bibnamefont {Chuah}},
  \bibinfo {author} {\bibfnamefont {R.}~\bibnamefont {Cazan}}, \bibinfo
  {author} {\bibfnamefont {M.~D.}\ \bibnamefont {Barrett}}, \ and\ \bibinfo
  {author} {\bibfnamefont {B.~K.}\ \bibnamefont {Sahoo}},\ }\href {\doibase
  10.1103/PhysRevA.88.012518} {\bibfield  {journal} {\bibinfo  {journal} {Phys.
  Rev. A}\ }\textbf {\bibinfo {volume} {88}},\ \bibinfo {pages} {012518}
  (\bibinfo {year} {2013})}\BibitemShut {NoStop}%
\end{thebibliography}
\end{document}